\documentclass[useAMS,usenatbib]{mn2e}
\usepackage{graphicx}

\title[Chemical evolution during gas-rich galaxy interactions]{Chemical evolution during gas-rich galaxy interactions}
\author[Perez et al.]{Josefa Perez$^{1,2,3,}$, Leo Michel-Dansac$^{4}$ and Patricia Tissera$^{1,3}$\\
$^{1}$Instituto de Astronom\'\i a y F\'\i sica del Espacio,Conicet-UBA, CC67, Suc.28,Ciudad de  Buenos Aires, Argentina.\\
$^{2}$C\'atedra de Astronom\'\i a General, Facultad de Ciencias Astronom\'\i a y Geof\'\i sica, Universidad Nacional de La Plata, Argentina.\\
$^{3}$Consejo Nacional de Investigaciones Cient\'\i ficas y T\'ecnicas, CONICET, Argentina.\\
$^{4}$Centre de Recherche Astrophysique de Lyon, Universit\'e de Lyon,
  Universit\'e Lyon 1, Observatoire de Lyon, \\ Ecole Normale Sup\'erieure de
  Lyon, CNRS, UMR 5574, 9 avenue Charles Andr\'e, Saint-Genis Laval, 69230,
  France.\\
}

\begin{document}

\date{Accepted ????. Received ????; in original form ????}
\pagerange{\pageref{firstpage}--\pageref{lastpage}} \pubyear{}
\maketitle
\label{firstpage}

\begin{abstract}
We analyse a set of galaxy interactions performed by using a
self-consistent chemo-hydrodynamical model which includes star
formation, Supernova feedback and chemical evolution.  In agreement
with previous works, we find that tidally-induced low-metallicity gas
inflows dilute the central oxygen abundance and contribute to the
flattening of the metallicity gradients. The tidally-induced inflows
trigger starbursts which increase the impact of SN II feedback
injecting new chemical elements and driving galactic winds which
modulate the metallicity distribution.  Although $\alpha$-enhancement
in the central regions is detected as a result of the induced
starbursts in agreement with previous works, our simulations suggest
that this parameter can only provide a timing of the first pericentre
mainly for non-retrograde encounters.  In order to reproduce wet major
mergers at low and high redshifts, we have run simulations with
respectively 20 and 50 percent of the disc in form of gas.  We find
that the more gas-rich encounters behave similarly to the less rich
ones, between the first and second pericentre where low-metallicity
gas inflows are triggered.  However, the higher strength of the
inflows triggered in the more gas-rich interactions produces larger
metal dilutions factors which are afterward modulated by the new
chemical production by Supernova.  We find that the more gas-rich
interaction develops violent and clumpy star formation triggered by
local instabilities all over the disc before the first pericentre, so
that if these galaxies were observed at these early stages where no
important tidally-induced inflows have been able to develop yet, they
would tend to show an excess of oxygen.  We find a global mean
correlation of both the central abundances and the gradients with the
strength of the star formation activity. However, the correlations are
affected by orbital parameters, gas inflows and outflows, suggesting
that it might be difficult to determine it from observations.
Overall, our findings show that a consistent description of the gas
dynamics and stellar evolution along the interactions is necessary to
assess their effects on the chemical properties of the interstellar
medium.
\end{abstract}

\begin{keywords}
galaxies: formation, galaxies: evolution, galaxies: interactions, galaxies: abundances.
\end{keywords}

\section{Introduction}

Galaxy interactions are considered a key process in galaxy formation
since these violent events could be able to redistribute mass and
angular momentum very efficiently
\citep[e.g.][]{BH96,tissera00,dimatteo08}.  As a consequence, chemical
elements can be stirred up affecting their distribution and
metallicity patterns such as gradients.  There are numerous
observational works dedicated to the study of galaxy pairs
\citep[e.g.][]{SP67,TL78,BGK00,LTAC03,NCA04,ALTC06,woods06,ellison08,ellison10,Patton11}.
But only in the last few years, it has been possible to focus on the
effects of galaxy interactions on chemical abundances.  Interacting
galaxies, galaxy pairs, ultra-luminous infrared galaxies, and merger
remnants have been found to exhibit low gas-phase metallicity at a
given stellar mass compared with the mean mass-metallicity relation
\citep[e.g.][]{KGB06,rupke08,ellison08,LMD08,peeples09,alonso10}.
Recently, \citet{kewley10} analysed the metallicity gradients of
galaxy pairs in detail confirming the central dilution and gradient
flattening previously reported by several authors.  These trends have
been interpreted as a consequence of low-metallicity inflows triggered
by galaxy interactions. Indeed, these interacting galaxies are in
majority spirals which, in the nearby Universe, hold a gas-phase
metallicity gradient \citep[e.g.][]{zaritsky94,dutil99,moustakas10}.
The presence of metallicity gradients could explain the decrease of
the central abundances when low-metallicity inflows are triggered
during interactions.  There are also some galaxies in pairs with
abundances departing significantly from the mass-metallicity relation
such as those reported by \citet{LMD08,peeples09,alonso10}.  Hence,
there are still effects of galaxy interactions which demand further
understanding.

By using cosmological chemo-hydrodynamical simulations,
\citet{perez06} showed for the first time that the interacting
galaxies tend to have lower central metallicity when compared to
isolated galaxies of similar stellar mass.  These authors showed that
the decrease of central abundances could be explained in terms of gas
inflows triggered during the interactions.  More recently other
numerical works have confirmed these findings by using different
simulations and hypothesis \citep{rupke10a,montuori10}.  Particularly,
\citet{rupke10a} studied the evolution of metallicity profiles by
using hydrodynamical simulations to recreate the major interaction of
galaxies with a gas mass fraction similar to that used in our less
gas-rich simulations.  However, these authors did not consider new
star formation activity and the subsequent chemical enrichment of the
interstellar medium as we do in our work.

In this paper, we use a chemical model grafted onto \texttt{GADGET-2},
which allows the description of the chemical enrichment of baryons as
the merger evolves.  Our code also includes a self-consistent model of
SN feedback \citep{scannapieco06}, so mass-loaded galactic outflows
can be triggered without including mass-depend parameters. This code
includes the chemical model of \citet{mosconi01} which describes the
enrichment of baryons by Type Ia and Type II SNe. Hence, the
simulations analysed in this work follow the dynamical evolution
together with the chemical evolution of baryons as the interaction
takes place.

This paper is organized as follows. In Section 2, we describe the code
and the initial conditions designed for this work.  In Section 3, we
explore wet interactions at low redshift while Section 4 analyses a
more gas-rich encounter, in order to reproduce typical wet
interactions at high redshift.  Finally, we summarize our findings in
Section 5.

\section{Numerical simulations}

We performed a set of hydrodynamical simulations of interacting
pre-prepared disc galaxies of comparable masses. The simulations were
run by using a version of \texttt{GADGET-2} \citep{springel05}
modified by \citet{scannapieco05, scannapieco06} to include the
chemical model of \citet{mosconi01} and a new SN feedback model
developed to provide an improved  description of the interstellar
medium (ISM) and the injection of energy and metals generated by SNe.
The multiphase and SN feedback model developed by \citet{scannapieco05, scannapieco06}
allow the coexistence of gas clouds with different thermodynamical properties and
are able to describe the injection of energy from young stars into the ISM producing the
self-regulation of the star formation activity as well as triggering  galactic outflows
without the need to introduce mass-dependent parameters or 
to change discontinuously particle momentum to start a wind 
or temporary suppression of radiative cooling.  
This scheme has  been successful at reproducing the star formation
activity during quiescent and starburst phases and is able to trigger mass-loaded winds 
with strength dependent on the potential well of the systems
\citep[e.g.][]{scannapieco06, derossi2010, sawala2011, scannapieco11}.

Chemical elements are synthesized in the stellar interiors, and
ejected into the ISM by SNe. The chemical model describes the
enrichment by SNII and SNIa, providing information on certain
individual chemical elements.  We also assume that each SN injects an
amount of energy into the ISM, $E_{\rm SN}$ (cf Table~\ref{tab1}).
SNII and SNIa are originated from different progenitors with different
rates, yields and time-scales. We assume that SNII are generated by
stars more massive than 8 M$_{\odot}$ after $\approx 10^6$ yrs. We
adopt the yields given by \citet{ww95}.  For SNIa the chemical model
assumes lifetimes randomly distributed in the range $[10^8, 10^9]$ yrs
and yields from W7 model of \citet{thielemann93}.  The SN feedback
model is able to generate galactic outflows powerful enough to
transport enriched material outside the galaxies
\citep{scannapieco08}.  Thus, these simulations provide us with the
evolution of the dynamical and the chemical properties of the baryons
as galaxy interactions proceed.

We run three major merger simulations (1:1) varying the orbital
parameters (SimI, SimII, and SimV) with the same star formation and SN
feedback model. We also performed a run (SimIV) with the same initial
conditions as SimII but decreasing the energy release by each SN event
(E$_{SN}$) and a run (SimIII) with the same orbital parameters but
without star formation and feedback activity. This last experiment is
comparable to those analysed by \citet{rupke10a}.  In order to
represent wet mergers at low redshift, these experiments (from SimI to
SimV) were run with a 20 percent of the disc in form of gas.  Finally,
SimVI has the same orbital, star formation and feedback parameters as
SimII but the gas fraction in the disc has been increased to up 50 per
cent in order to simulate a wet encounter at high redshift.  The
parameters of the simulations are summarized in Table ~\ref{tab1}.

Initial conditions correspond to disc galaxies composed of a DM halo,
a bulge component (modeled by a NFW and a Hernquist profiles,
respectively) and an exponential disc, with a total baryonic mass of
$M_b \sim 5 \times 10^{10} M_{\odot}$. For the less gas-rich
simulations, we use $200\,000$ dark matter particles, $100\,000$ stars
distributed in the stellar disc and the bulge, and $100\,000$
particles to represent the disc gas component with an initial gas
resolution of $\approx 3 \times 10^5$ M$\odot$.  For the more gas-rich
experiment, we re-calculate the number of particles so as to keep
similar mass resolution.  Total stellar masses change according to the
gas content in the galaxy discs, being $M_* \sim 4.4 \times 10^{10}
M_{\odot}$ for the less gas-rich experiment and $M_* \sim 3.2 \times
10^{10} M_{\odot}$ for the more gas-rich simulation.  A gravitational
softening of $\epsilon_{\rm G}=0.16$ kpc was adopted for the gas
particles, $\epsilon_{\rm S}=0.2$ kpc for the stars, and
$\epsilon_{\rm DM}=0.32$ kpc for the dark matter. Galaxies live in DM
halos with circular velocities of $160 \, {\rm km/s}$.  All these
experiments are set to have the same elliptical orbits, according to
\citet{TT72}, with pericentre distance of $\approx 20 $ kpc and
circularity parameter $\epsilon \sim 0.20$.  These orbital values are
consistent with those obtained from the analysis of cosmological
simulations \citep[e.g.][]{kb06}. We have kept them fixed so that
changes in the chemical and dynamical properties can only be adscribed
to the different angular momentum orientation.  A larger orbital
distribution might be of interest to analyse possible dependences in
the future.

The gas component is initially assumed to have been pre-enriched by
the existing stars, following a metallicity gradient statistically
consistent with observations.  We assigned an initial amount of
chemical elements to the gas particles in order to reproduce the mean
observational gradient reported by \citet{dutil99}, which for the less
gas-rich simulations is $-0.08$ dex kpc$^{-1}$ with a central oxygen
abundance of 9.2.  Thus, the initially simulated galaxies are located
onto the observed mass-metallicity relation \citep{tremonti04}.  For
the more gas-rich simulation (SimVI), the initial metallicity gradient
is somewhat steeper ($-0.1$ dex kpc$^{-1}$) and the central
metallicity lower (by 0.7 dex) than the less gas-rich galaxies. By
doing that, these more gas-rich simulated galaxies are then initially
onto the mass-metallicity relation observed at high redshift by
\citet{maiolino08}.  Note that the exact adopted values are not
critical since we are interested in the relative evolution of the
metallicity.

\begin{table}
  \begin{center}
    \caption{Characteristics of the numerical experiments.  $i$: angle
      between the two total angular momentum of the
      galaxies. E$_{SN}$: amount of SN energy release by each event.
      $f_{\rm gas}$: gas fraction of the discs. }
    \label{tab1}
    \begin{tabular}{|l|c|c|c}\hline
       {Simulation}   & $i$ ($\degr$) & E$_{SN}$ ($10^{51}$ erg/s) & $f_{\rm gas}$\\
      \hline
       SimI    &  $45$           & $0.5$ & 0.20 \\
       SimII   &  $0$           & $0.5$ & 0.20 \\
       SimIII (no SF)  &  $0$           & $0.$  & 0.20 \\
       SimIV   &  $0$           & $0.04$ & 0.20 \\
       SimV    &  $180$         & $0.5$ & 0.20\\
       SimVI &  $0$          & $0.5$ & 0.50 \\
      \hline
    \end{tabular} 
  \end{center}
  \vspace{1mm}
\end{table}

\section{Metallicity evolution of wet mergers in the local Universe}\label{dinamica}

\begin{figure}
  \centering
  \includegraphics[width=84mm]{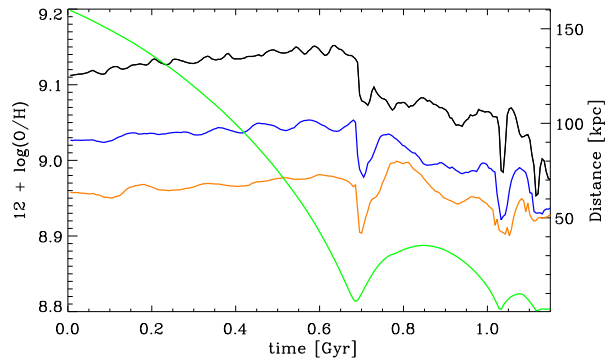}
  \caption{Evolution of the gas-phase oxygen abundance measured inside
    a sphere of 2 kpc (black line), 4 kpc (blue line) and 6 kpc
    (orange line) radius for one of the interacting galaxies of
    SimI. The relative distance between the center of mass of the
    simulated galaxies is also plotted (green line). }
  \label{leo1}
\end{figure}

In this section, we analyse the effects of galaxy interactions on the
central metallicity in wet local encounters. Taking SimI as an example,
we show in Fig.~\ref{leo1} the evolution of the gas-phase O/H
abundance of one member of the galaxy pair\footnote{The other member
  of the pair have a similar behaviour.}  as the interaction
proceeds. The evolution of the interaction can be followed by the
relative distances between the center of masses of the interacting
systems, plotted in the same figure.  As it can be seen, there is a
clear evidence of a decrease in the central gas-phase metallicity
which can be correlated with the first and the second pericentre.
Several previous works have shown that during a close encounter,
strong tidal torques can develop, re-distributing angular momentum and
producing gas infall \citep[e.g.][]{BH96,MH96,tissera00,dimatteo08}.In
the case that a metallicity gradient exists, the infalling gas will
tend to have lower metallicity than that of baryons in the central
region producing a dilution as it is clearly shown by our simulation
\citep[see also][]{rupke10a,montuori10}.

In order to quantify the chemical changes as a function of time, we
compute the central abundances inside concentric spheres of different
radius in the range from 1 kpc to 10 kpc.  As an example,
Fig.~\ref{leo1} shows the gas-phase oxygen abundance measured inside a
sphere of 2 kpc, 4 kpc, and 6 kpc of radius.  As it can be
appreciated, as one moves inwards, the signals get noisier since the
number of particle decreases. If one moves to larger radius, the
signal decreases as chemical abundances start to be dominated by the
external regions. As a compromised between these two effects, we
select $R_{\rm cen}=4$ kpc (blue line) to define the central region
where we will perform the analysis of central metallicities.

\begin{figure}
  \centering
  \includegraphics[width=84mm]{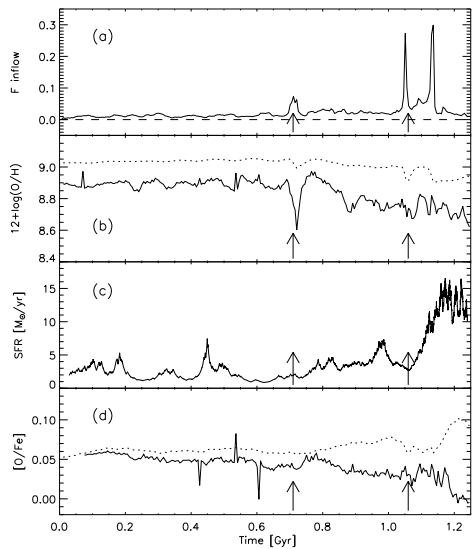}
  \caption{Evolution of gas inflow and of chemical abundances in the
    central region $R < R_{cen}$ of one galaxy in SimI . a) Fraction
    of the gas falling within $R_{cen}$ as a function of time. Dashed
    horizontal line represents the initial fraction. b) Oxygen
    abundance of the infalling gas (solid) and the mean central
    abundance within $R_{cen}$ (dotted line).  c) Evolution of star
    formation rate. d) $\alpha$-enhancement of the infalling gas
    (solid) and its mean central value (dotted line). Arrows indicate
    the first and second pericentres.}
  \label{general}
\end{figure}

In order to understand the decrease in the central abundances 
 shown by Fig.~\ref{leo1}, we analyse the
evolution of the central gas inflows and their abundances.  We compute
the central gas inflow at a certain time as the ratio between the new
accreted gas mass within $R_{\rm cen}$ and the total gas content
within the same radius.  Fig.~\ref{general}a shows the evolution of
this central gas inflow as the interacting galaxies approach each
other.  We find that the gas inflow remains constant in around $2\%$
until the first close passage when it rises up to $\sim 8\%$.  From
the second pericentre, the gas inflow steadily increases up to around
$30\%$ .  This gas inflow is dominated by low-metallicity material as
it can be appreciated from Fig.~\ref{general}b (solid line) where we
plot its O/H abundance.  The accretion of this low-metallicity gas
dilutes the central oxygen abundance as shown by the continuous
decrease detected after the first pericentre (Fig.~\ref{general}b,
dotted line).  The main drops in central abundances correlate with the
occurrence of the pericentres driven by important gas inflows
(Fig.~\ref{general}a).  Additionally, the gas inflow is also
responsible for triggering new star formation activity
(Fig.~\ref{general}c).  In SimI, the SFR increases by a factor 3 from
the first pericentre as the interaction continues fueling the gas
inflow (Fig.~\ref{general}a,c).  As a consequence of this increase in
the SFR, the enrichment by SNII also increases, diminishing the
effects of chemical dilution previously imprinted by the
low-metallicity gas inflows.

The increase of the SFR is also reflected in the $\alpha$-enhancement,
[O/Fe], of the gas in the central region (dotted line of
Fig.~\ref{general}d).  This central $\alpha$-enhancement is present
from the first pericentre but is much stronger around the second one.
Note that the infalling gas into the central region is low
$\alpha$-enhanced as expected in a disc galaxy such as that of the
Milky Way (solid line of Fig.~\ref{general}d), indicating that the
increase of oxygen in the central region is principally due to SNII
associated to the central starburst.  These general results are in
agreement with those reported by \citet{montuori10} although these
authors did not consider chemical enrichment self-consistently. As we
will discuss in the next section, our simulations allow us to make further
advances on this issue.

Previous works \citep[e.g.][]{BH96,dimatteo08,cox08,hopkins09} have
already reported how star formation histories of interacting systems
depend on the initial conditions and orbital parameters.  Hence, the
trends shown in Fig.~\ref{general} for SimI might vary among the
analysed simulations as we will discuss in next sections.

\subsection{Could the $\alpha$-enhancement clock the interaction?}

As already shown for SimI, the $\alpha$-enhancement starts to be
detected after the first close passage. However, the most significant
increase is recorded after the second one. In order to analyse if the
$\alpha$-enhancement could be used to time the interaction as
suggested by \citet{montuori10}, we study the ISM enrichment history
along the interactions and its dependence on different orbital
configurations.

\begin{figure}
  \centering
  \includegraphics[width=84mm]{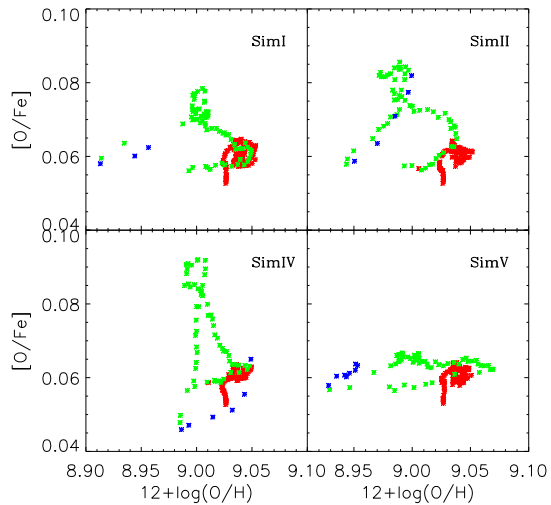}
  \caption{Evolution of central abundances as galaxy interaction
    proceeds for different simulations: I, II, IV and V.  Red symbols
    identify the phase of interaction before the first pericentre,
    green points the phase between the first and second pericentres
    and blue ones the final merging phase after the second
    pericentre.}
  \label{alpha1}
\end{figure}

Fig.~\ref{alpha1} shows the evolution of [O/Fe] in the central regions
for different simulations with different orbital parameters: SimI,
SimII and SimV.  We also include SimIV which assumes a lower energy
per SN event but otherwise has the same parameters as SimII.  We find
that before the first close passage (red points), the different
simulations exhibit no significant variations in the $12 + \log$(O/H)
nor [O/Fe] values.  However, different feedback parameters and orbital
configurations impact differently on the ISM chemical enrichment
during the interacting phase between the first two pericentres (green
points).  As a general result, we find an increase of the [O/Fe],
except for the coplanar retrograde orbit (SimV).

\begin{figure}
  \centering
  \includegraphics[width=84mm]{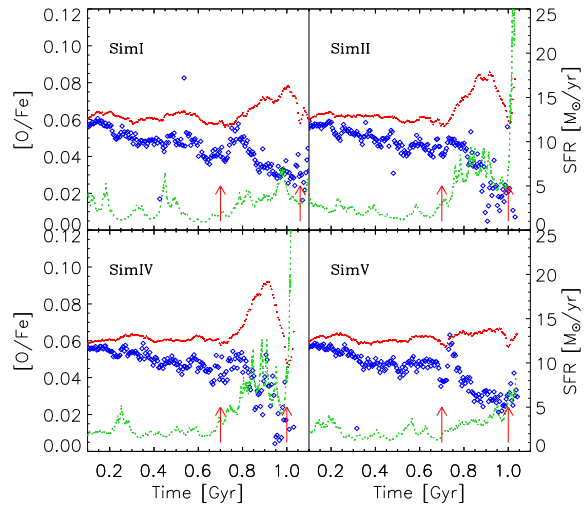}
  \caption{Evolution of $\alpha$-enhancement of the central region
    (red dotted lines) and of the infalling gas (blue open symbols)
    and the star formation rate (green lines) for different
    simulations: I, II, IV and V. Arrows in each panel indicate the
    first and second pericentres.}
  \label{alpha3}
\end{figure}

In order to understand these differences in the $\alpha$-enhancement,
Fig. \ref{alpha3} shows the central $\alpha$-enhancement and the SFR
for one of the interacting galaxies in each simulation. As expected,
the central $\alpha$-enhancement (red dotted lines) correlates with
the star formation activity (green lines).  We find that between the
first and the second pericentre the [O/Fe] reaches the highest values
for coplanar direct orbits (SimII and SimIV), decreases as inclination
increases (SimI), and  is almost negligible for the coplanar
retrograde interaction (SimV).  Note that in agreement with previous
results by \citet{dimatteo07}, we find that coplanar direct
interactions trigger stronger starburts than coplanar retrograde
configurations during the pre-merger phase (see Fig. 8 of their
paper). 
We also find that the simulation with standard feedback (SimII) has
slightly lower levels of the [O/Fe] and SFR, than the case of low
energy feedback (SimIV) because in the later, SN feedback drives
weaker galactic outflows, being less efficient at transporting
material out of the central regions.

As discussed previously by \citet{montuori10}, the
$\alpha$-enhancement observed in interacting galaxies before the final
coalescence of systems might date the interaction. However, we find
that this is valid when the $\alpha$-enhancement is only regulated by
the star formation process. In this case, the increase of [O/Fe]
during the interacting phase would indicate the occurrence of a recent
starburst (approximately $<$ 1 Gyr), because the ISM have not had
enough time to be heavily enriched by SNIa yet.  On the contrary, the
detection of a decreasing [O/Fe] value would indicate that SNIa had
time to release Fe into the ISM, which means that the time elapsed 
since the last starburst is longer than approximately $\sim 1$ Gyr.
Our simulations which follow the dynamical and chemical evolution
along the interaction suggest that other factors are also involved in
the regulation of the $\alpha$-enhancement.  As can be seen for SimI,
SimII and SimIV (Fig.~\ref{alpha3}), the increase in the
$\alpha$-enhancement after the first pericentre is followed by a drop
before the second pericentre.  We find that this dilution of the
central $\alpha$-enhancement is caused by the low [O/Fe] enhancement
of the very important gas inflow (open blue symbols in
Fig.~\ref{alpha3}), triggered during the second close approach of the
galaxies.  This gas inflow vanish the clear traces in the central
abundances left by the star formation activity, restarting the clock
to date the interaction.

Hence, our simulations suggest that the $\alpha$-enhancement can only
indicate the proximity to the first pericentre, providing extra
information on the parameters of the encounter since those with orbits
close to coplanar retrograde configurations would tend to have very
mild enhancement (SimV in Fig.~\ref{alpha3}).  However, our findings
also suggest that the chance to observe close galaxy pairs with
central $\alpha$-enhancement might be low, because of the short time
interval over which it could be detected (less than approximately 0.2
Gyr according to our experiments).

\subsection{Gas-phase metallicity profiles of wet low-redshift interactions}

\begin{figure*}
\includegraphics[width=6.09cm]{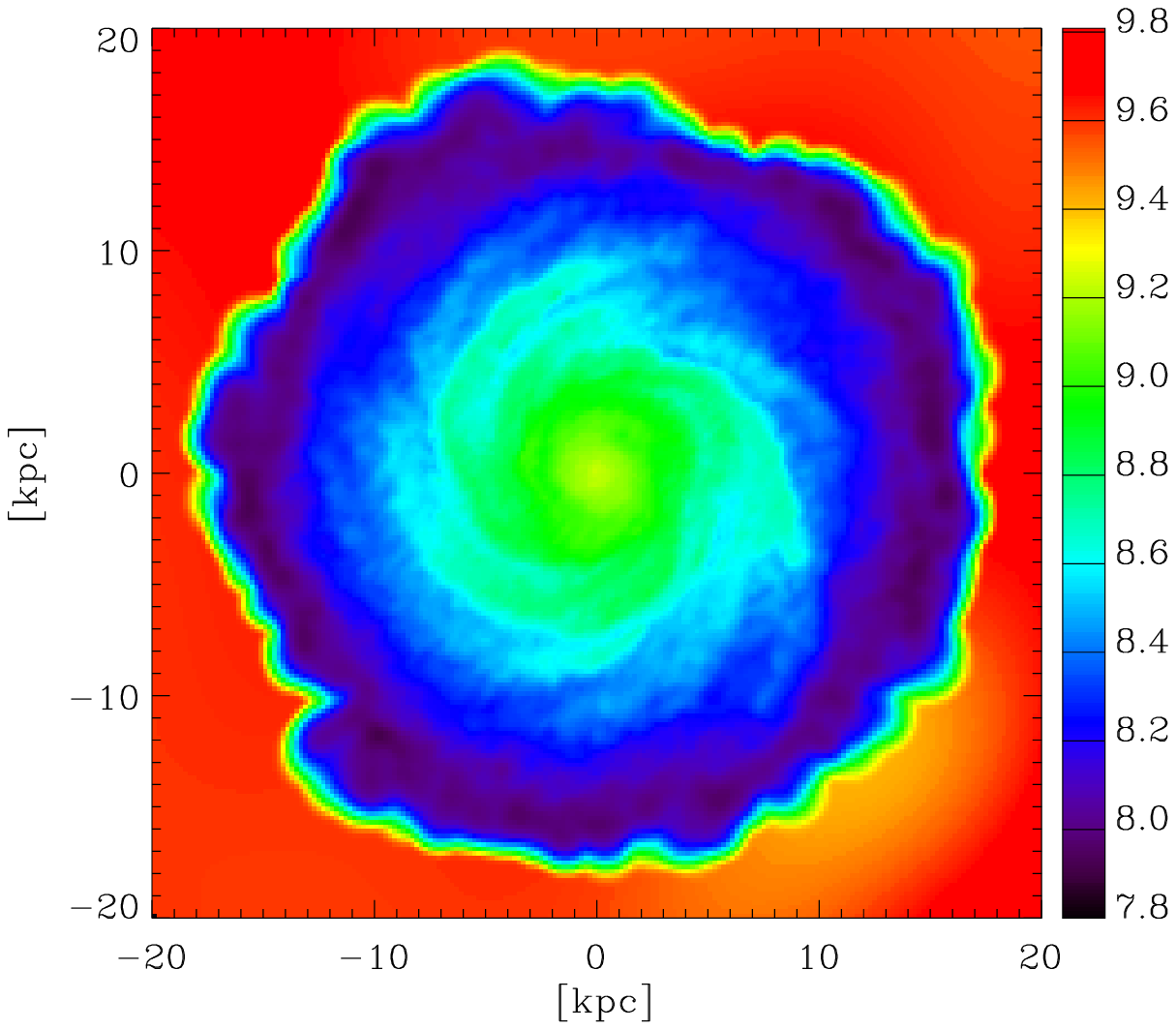}
\includegraphics[width=5.30cm]{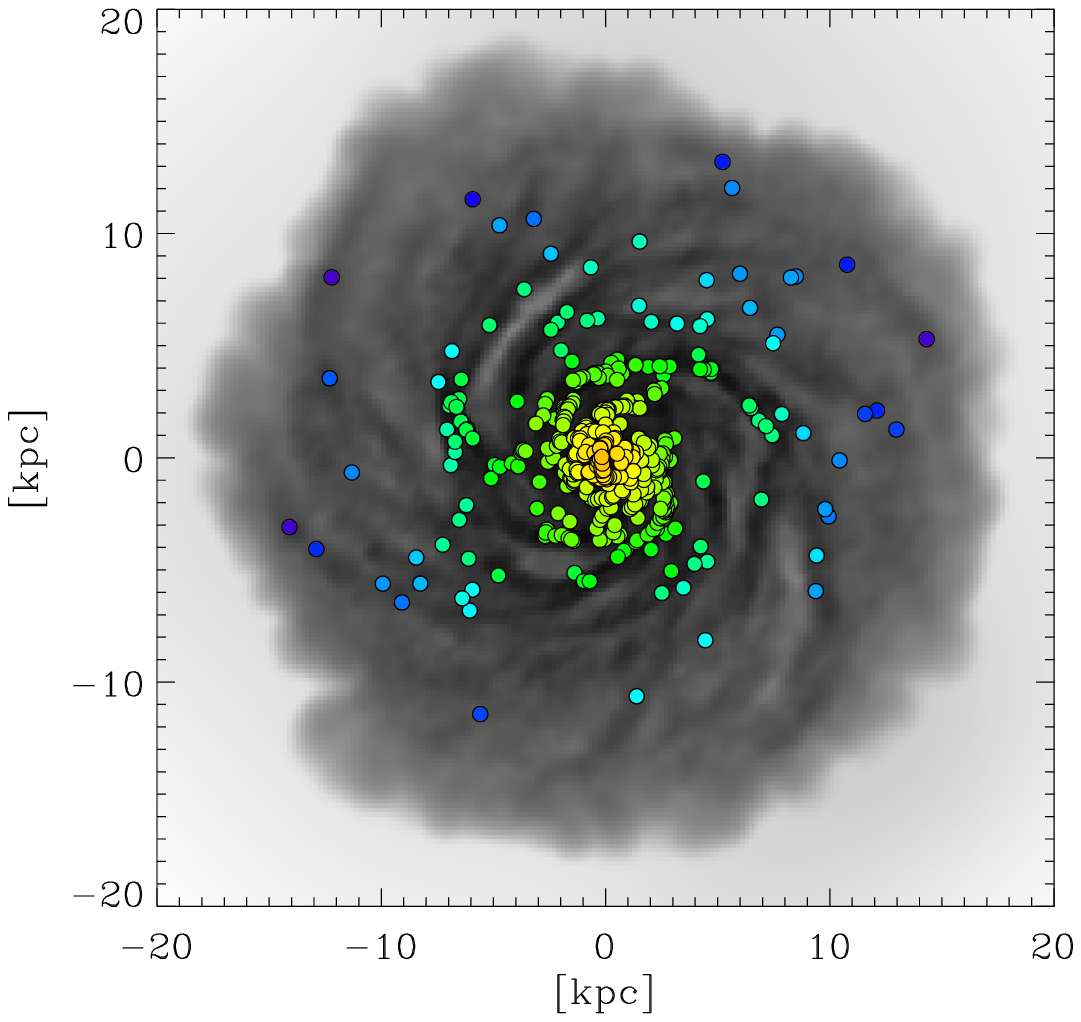}
\includegraphics[width=6.00cm]{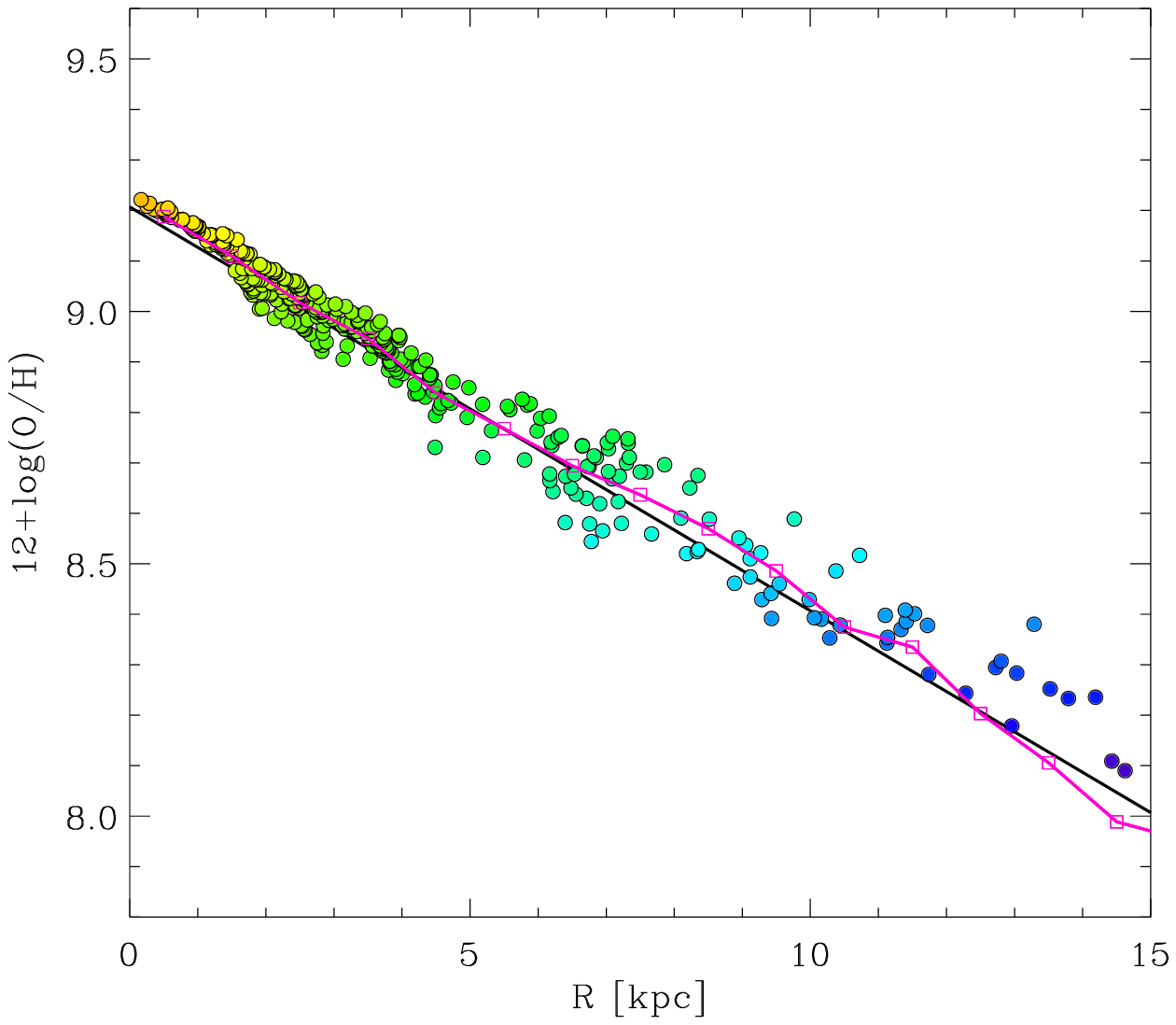}

\includegraphics[width=6.09cm]{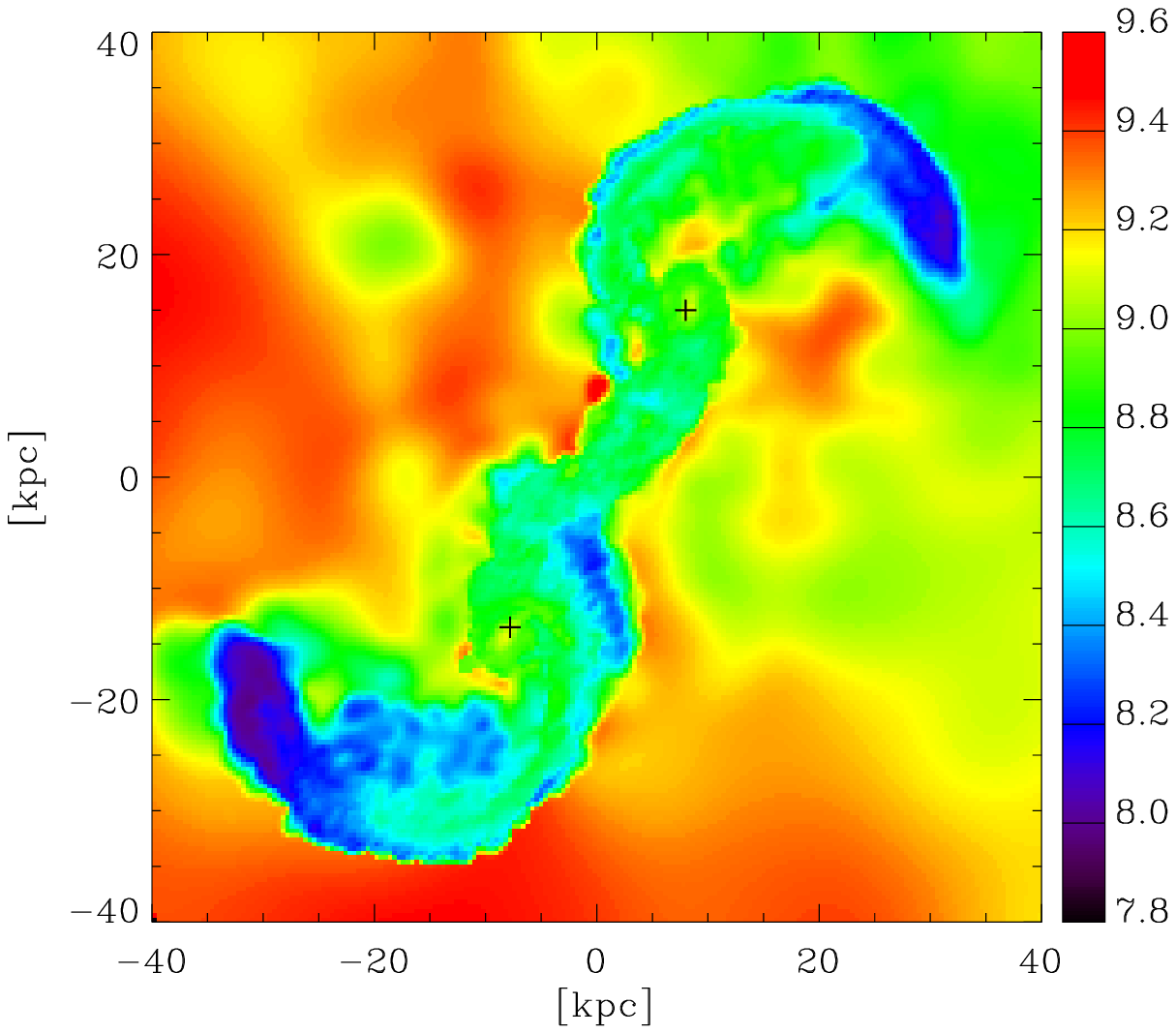}
\includegraphics[width=5.30cm]{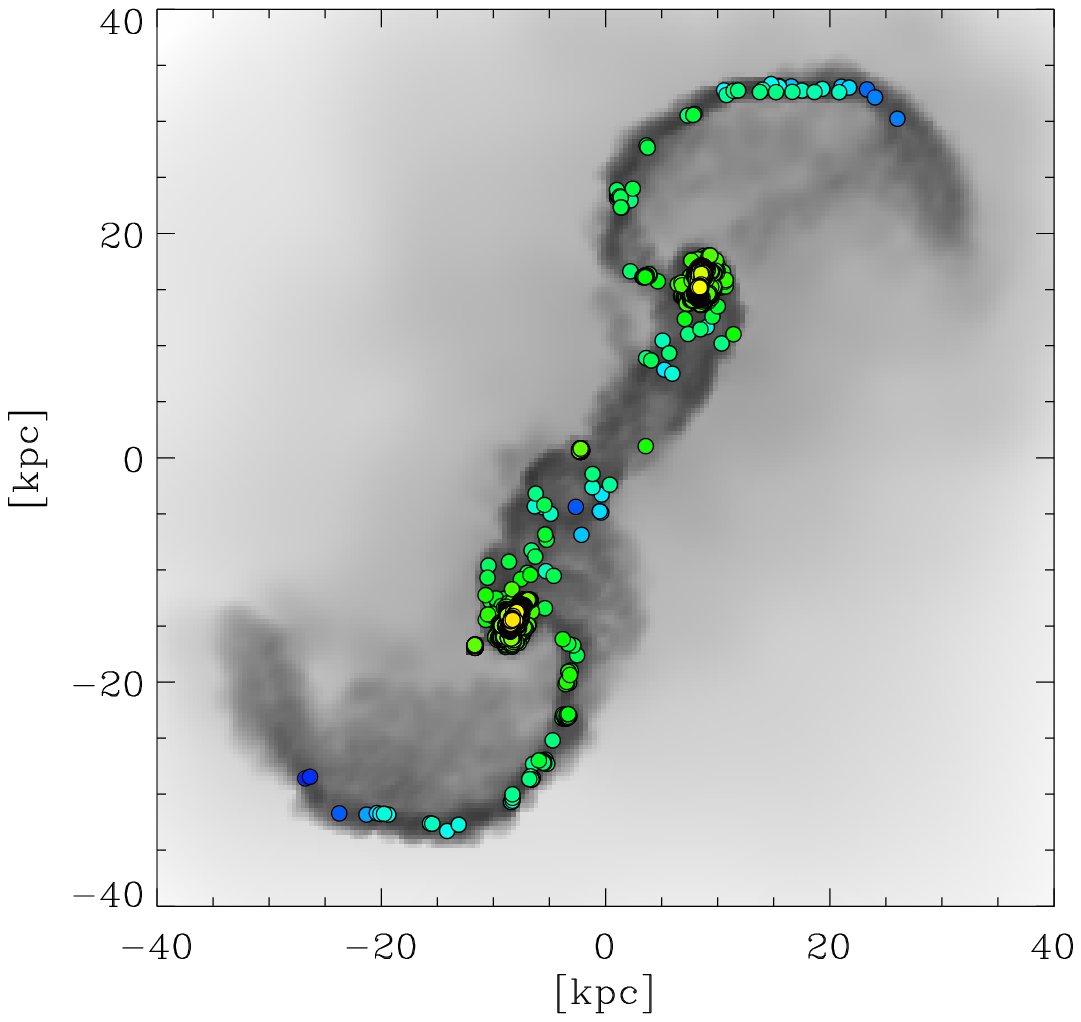}
\includegraphics[width=6.00cm]{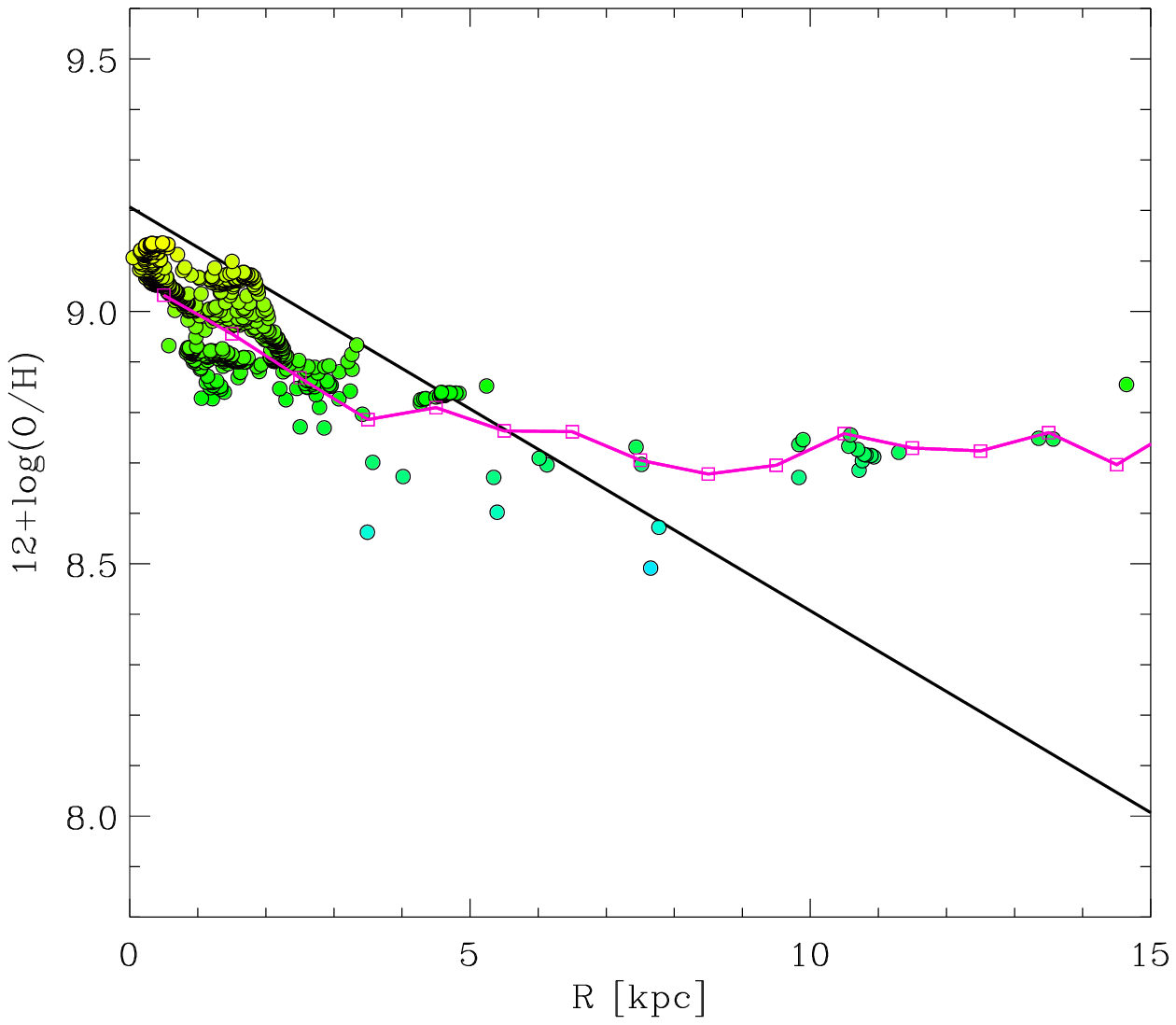}

\includegraphics[width=6.09cm]{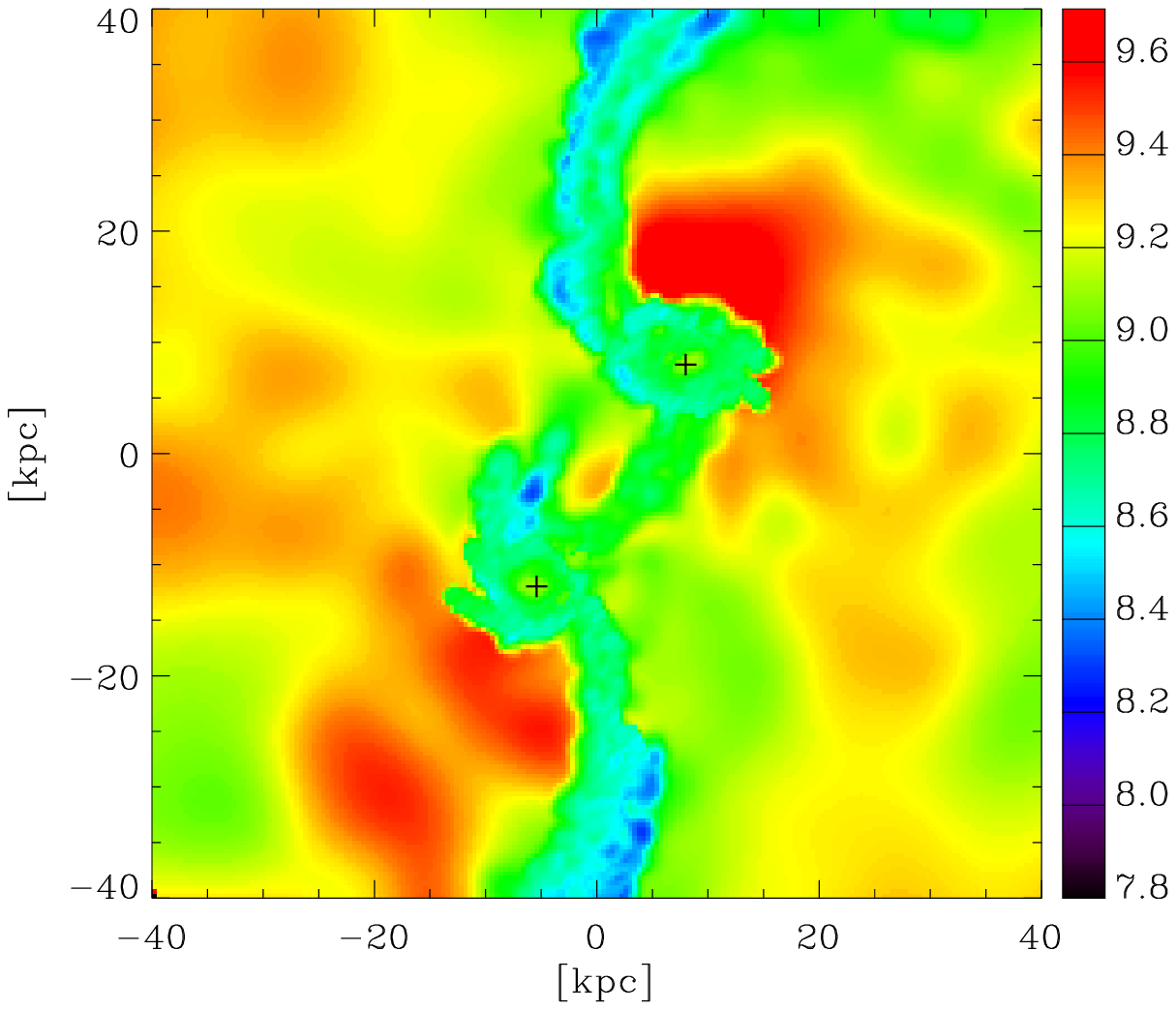}
\includegraphics[width=5.30cm]{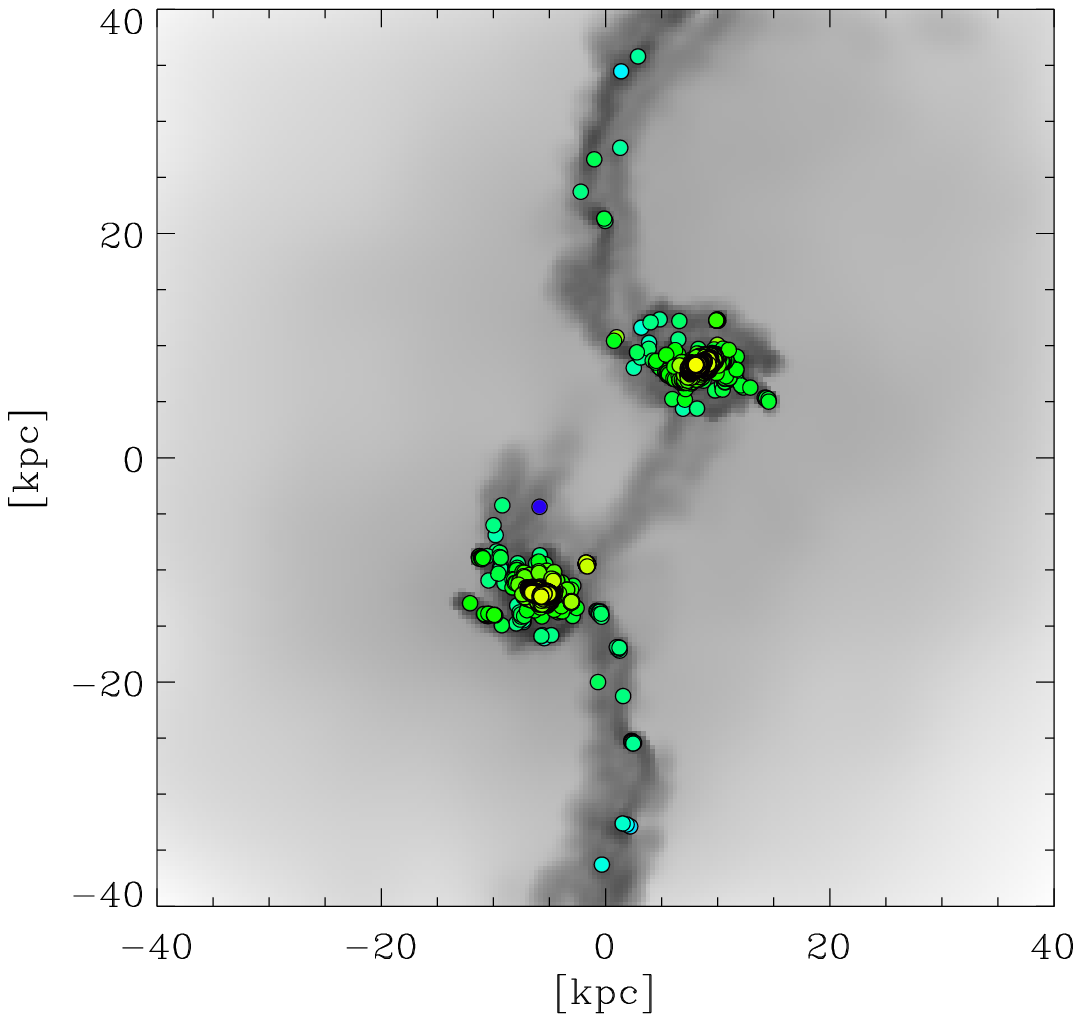}
\includegraphics[width=6.00cm]{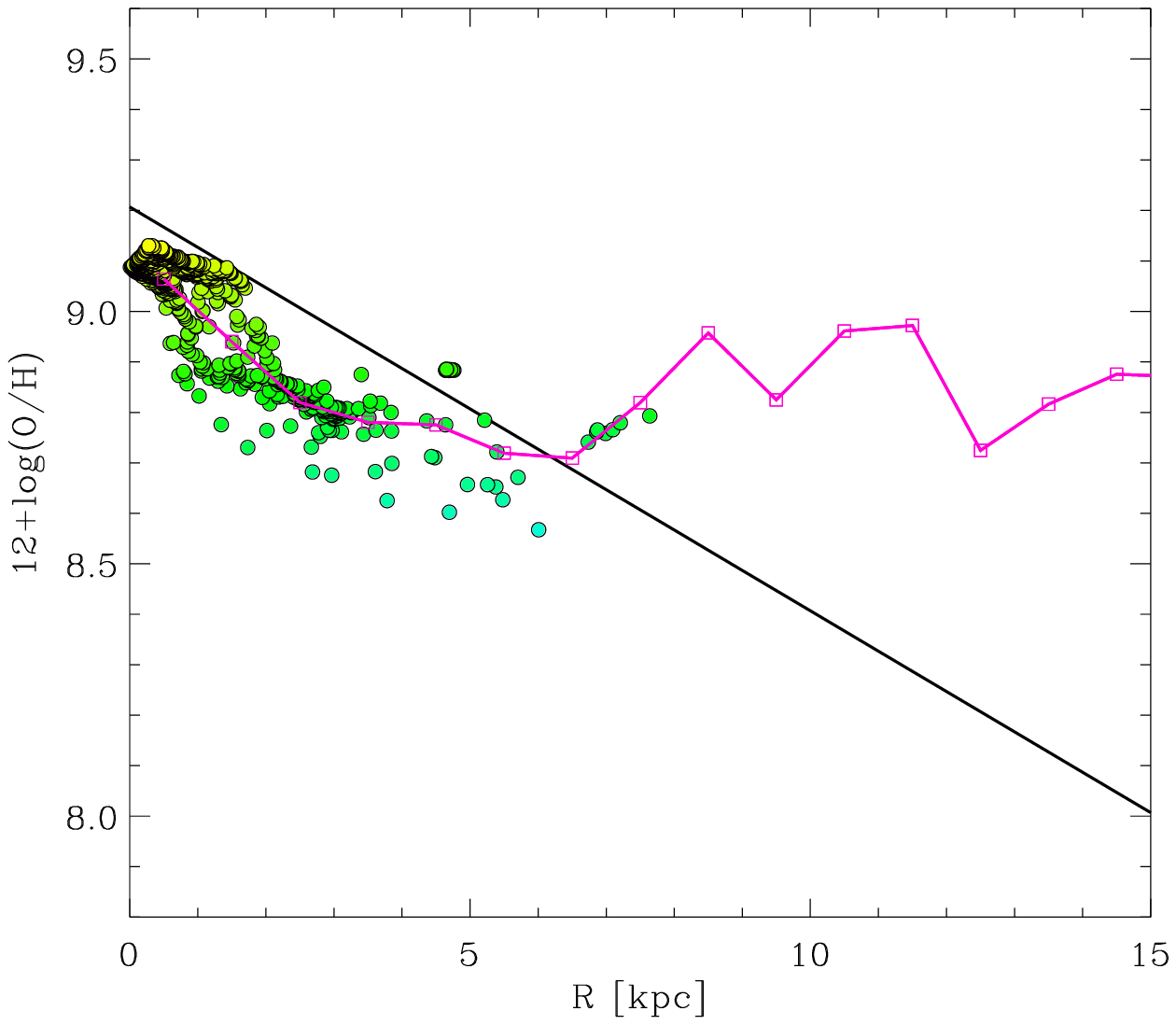}
\caption{Time sequence of the chemical properties in the gas component
  of SimII (before the first pericentre at $t=0.3$Gyr (upper panel),
  after the first pericentre at $t=0.8$Gyr (middle panel) and just
  before the second pericentre at $t=0.9$Gyr (lower panel)) . Left
  panels: oxygen abundance maps. Middle panels: gas density map (in
  grey) and oxygen abundances of the star-forming regions (colour
  circles colour-coded as the left panels). Right panels: gas-phase oxygen
  abundance profiles. The black line shows the initial metallicity
  gradient. The magenta line and symbols show the gas-phase
  metallicity gradient computed with all the gaseous particles. For
  the top panels, only one galaxy is shown since the two galaxies are
  very distant and the field of view is smaller than the bottom
  panels. The right panels show that the abundance gradients computed
  by radial binning of all the gaseous particles are in good
  agreements with the gradients determined with only the abundances of
  star-forming regions, as it is done with observations. }
\label{leo-map}
\end{figure*}

As shown in the previous section, galaxy-galaxy interactions have a
clear effect on the central abundances and the star formation of
wet interacting systems at low redshift caused by the dynamical perturbation
induced on the gas component.  In order to illustrate how this affect
the gas-phase metallicity profiles, we show a sequence of oxygen
abundance maps which plot the spatial distribution for the interacting
galaxies of SimI at different stage of the interaction, with their
corresponding oxygen abundance profiles (Fig.~\ref{leo-map}).  It can
be appreciated how the metallicity profile changes as the interacting
system gets closer.  This evolution is mainly related to the
flattening of the profile caused by a decrease of central
metallicities and a simultaneous increase of the abundances in the
external region of the disc.  We also show in Fig.~\ref{leo-map}
(right panel) that the abundance gradients computed by radial binning
of all the gaseous particles are in good agreements with the gradients
determined with only the abundances of star-forming regions, as it is
done with observations.

\subsubsection{Internal oxygen abundance profiles}

In order to quantify the evolution of internal metallicity profiles,
we carry out a linear regression fit to estimate the central oxygen
abundances, (O/H)$_{C}$, and slopes.  The internal profiles are
defined in the range of 3$\epsilon_{G}-7$ kpc. The inner radial limit
corresponds to three times the gravitational softening used for
baryons, and for the external radius, $R_{\rm ext}=7$ kpc corresponds
to approximately three scale-lengths of the initial stellar disc.

\begin{figure}
  \centering
  \includegraphics[width=84mm]{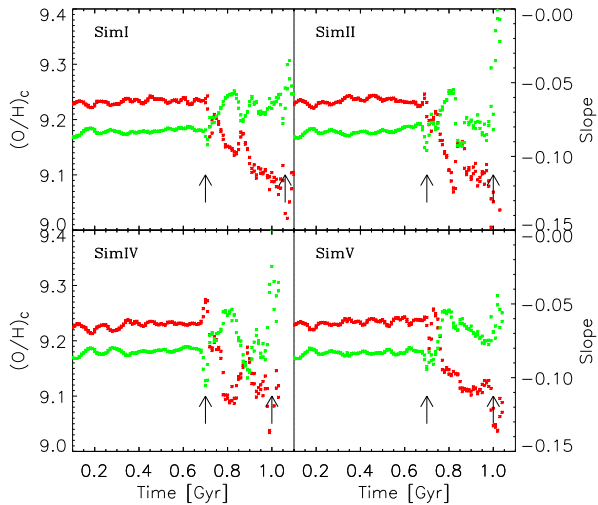}
  \caption{Evolution of the linear regression coefficients
    ((O/H)$_{C}$: red and Slope: green) of the oxygen abundance radial
    profile, computed in a range of 3$\epsilon_{G}-7$ kpc for
    different simulations: I, II, IV and V.  Arrows indicate the first
    and second pericentres.}
  \label{fit1}
\end{figure}

Fig.~\ref{fit1} shows the linear fitting coefficients of the radial
profiles for different simulations.  As it can be appreciated, there
is a significant evolution of internal profiles after the first
pericentre, which is consistent with a central dilution of abundances
as expected from the discussion of the previous subsection.  In
agreement with observations \citep{KGB06,rupke08} and previous
numerical works \citep{rupke10a}, the drop in (O/H)$_{C}$ is about 0.2
dex (red points).  We also find that the final metallicity gradients
are flatter than those in earlier stages of the interaction (green
points) in agreement with \citet[][and references
  therein]{kewley10,rupke10b}.

Note that although the evolution of internal profiles seems to be not
very sensitive to different orbital configurations in 
wet major mergers at low redshift, our results show some evidence that
galaxies in coplanar direct orbits (SimII and SimIV of
Fig.~\ref{fit1}) experience a larger flattening than those in coplanar
retrograde ones (SimV in Fig.~\ref{fit1}).

\subsubsection{External oxygen abundance profiles}

As mentioned above, the external abundance profiles rise as the galaxy
interaction proceeds. This is a common feature to all simulations,
hence, we will take SimI again for illustration purposes.

We define the external region of the gaseous disc as everything
outside $R_{\rm ext}= 7$ kpc, which also denotes quite well the radius
where the outer profiles depart from that defined by the central
region (Fig.~\ref{leo-map}).  Then, we estimate the fraction of gas
which has gotten into the external disc between two available time
intervals, as well as the metallicity of this gas as a function of
time.

\begin{figure}
  \centering
  \includegraphics[width=84mm]{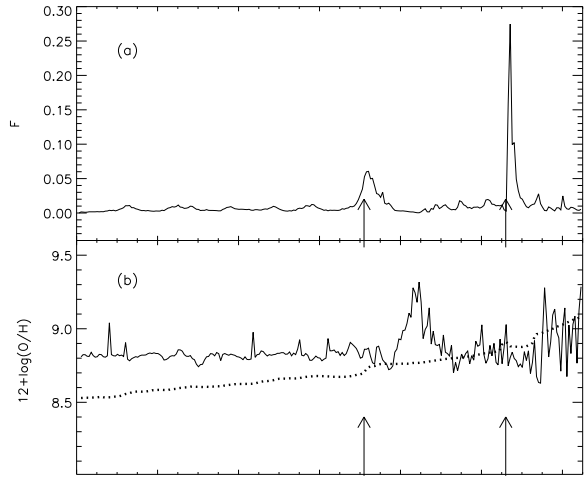}
  \caption{a) Fraction of gas getting into the external region ($R >
    R_{ext}=7$ kpc) as a function of time.  b) Oxygen abundance of the
    gas getting into this external region (solid line) and the mean
    external abundances (dotted line) as a function of time.  Arrows
    indicate the first and second pericentres. These calculations are
    shown for one galaxy member of SimI as an example.}
  \label{out}
\end{figure}

As can be seen from Fig.~\ref{out}a, close to each pericentre, the
external disc of interacting systems receives a significant fraction
of gas, principally close to the second one ($\sim 30\%$ at the second
pericentre).  This gas comes from the inner regions of the disc as the
arms get opened and distorted by tidal forces (Fig.~\ref{leo-map}).
As a consequence, this gas tends to have high oxygen
abundances (Fig.~\ref{out}b, solid line), which partially contribute,
after the first pericentre, to increase the mean external abundances
by $\approx 0.4$ dex (Fig.~\ref{out}b, dotted line).
This result is in agreement with \citet{rupke10b}, who find that tidal
tails transport metal-enriched gas from the galaxy inner regions to
the outskirts.

On the other hand, there is new star formation activity triggered in
the external regions, on the spiral arms as can be seen from
Fig.~\ref{leo-map}. These new stars will also contribute with new SN
II to the enrichment of the external ISM.  Therefore, the evolution of
the external abundances are produced by the combined effects of these
two processes: i) the {\it in situ} SF activity and ii) the gas-rich material
coming from inner regions as the interaction takes place and the spiral
arms get distorted.

\section{Metallicity evolution in wet high-redshift mergers}

In previous sections, we analyse wet low-redshift interactions of galaxies
with around 20 per cent of the disc in the form of gas. Here, we will
discuss the metallicity evolution in interacting systems when the
initial amount of gas is increased up to 50 per cent (SimVI). This
scenario is suitable for studying wet galaxy encounters at high
redshift.

For this simulation, we have chosen the same orbital configuration of
SimII, but with a modified initial distribution of baryons in order to
start with more gas-rich systems (cf. Table ~\ref{tab1}). We have also
adopted the same star formation and SN parameters as for SimII.  The
initial abundance gradient has been adjusted to be consistent with the
observations of the mass-metallicity relation at high redshift.
Particularly, we used the observations of \citet{maiolino08} at $z
\sim 3.5$.  Hence, the initial zero point is lower and the gradient
steeper than in the case of less gas-rich interactions (see Section 2).

\begin{figure}
  \centering
  \includegraphics[width=9.cm,height=11.5cm]{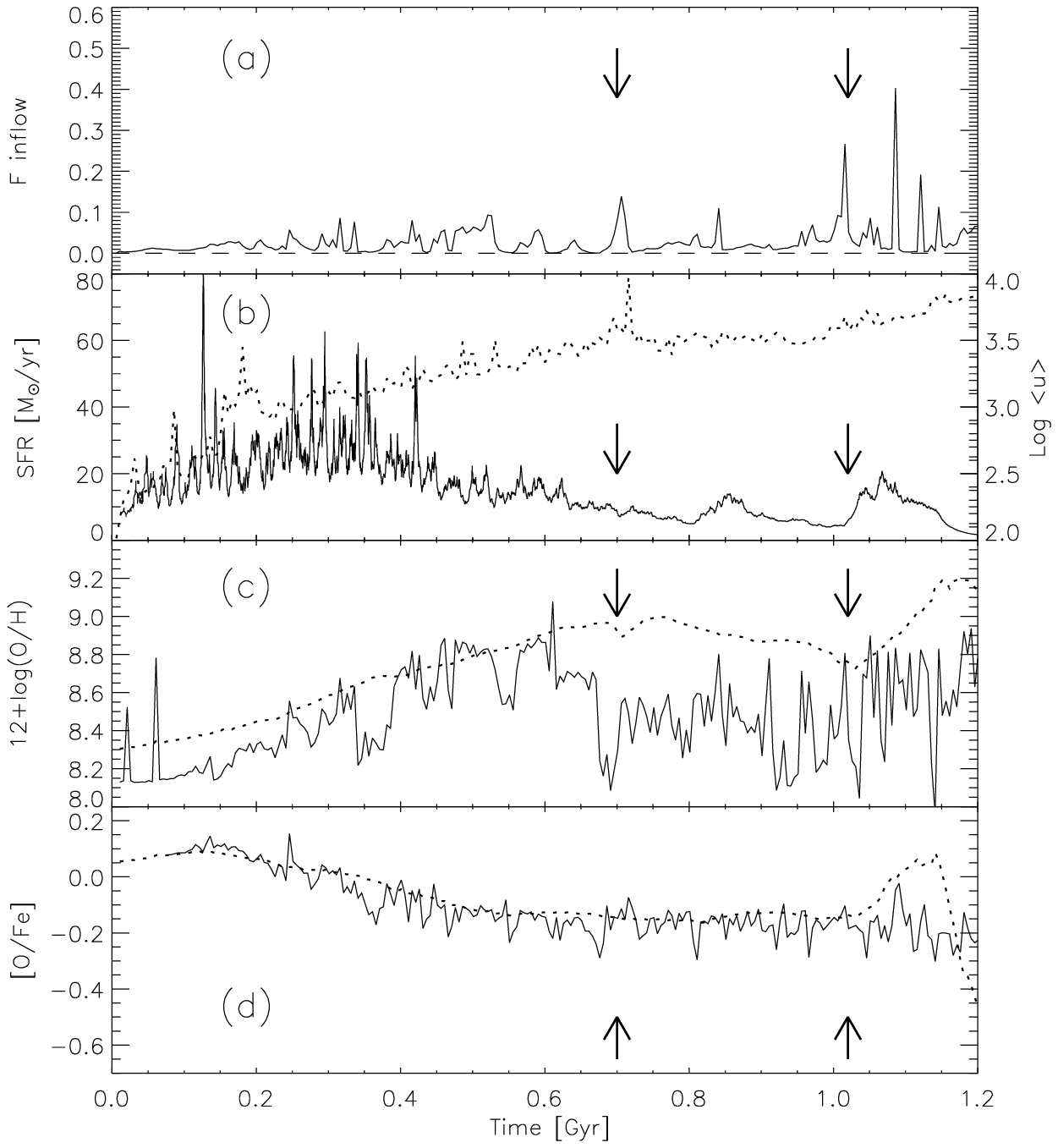}
  \caption{High gas-rich simulation (SimVI): a) Fraction of gas inflowing
    within $R_{\rm cen}$ as a function of time.  b) Evolution of star
    formation rate (solid line) and of the mean internal energy $<u>$
    of the gas (dotted line).  c) Oxygen abundance of the infalling
    gas (solid line) and the mean central oxygen abundance (dotted
    line), computed within $R_{\rm cen}$.  d) $\alpha$-enhancement of
    the infalling gas (solid line) and it mean central value (dotted
    line).  Arrows indicate the location of first and second
    pericentres.}
  \label{gasrich1}
\end{figure}

Fig.~\ref{gasrich1}a shows the gas inflows for one of the galaxy
members in SimVI. In this case, the amount of gas that falls into the
central region is more significant than in its less gas-rich
counterpart throughout the whole simulated time interval. Particularly
after the second pericentre, the gas inflow roughly duplicates that of
the less gas-rich interacting galaxies.

\begin{figure*}
  \centering
  \includegraphics[width=17.5cm]{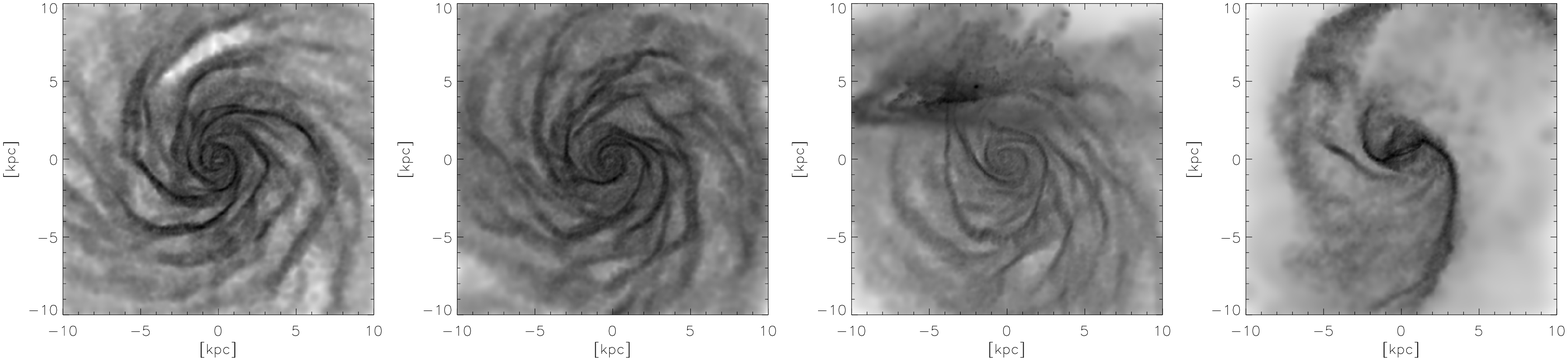}\\
  \includegraphics[width=17.5cm]{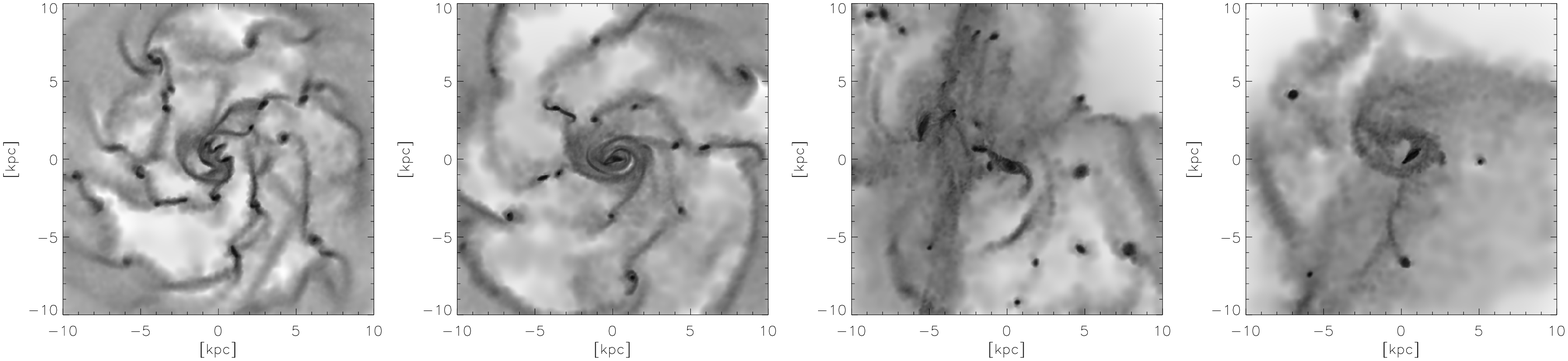}
  \caption{Projected gas mass distribution onto the disc plane of one
    interacting galaxy for the less gas-rich simulation SimI (upper
    panels) and for its more gas-rich counterpart SimVI (bottom
    panels). Gas density maps are shown from the left to the right, at
    0.3 Gyr when the galaxies are separated by $\approx $100 kpc, well
    before the first pericentre; at 0.45 Gyr; just at the first
    pericentre at 0.7 Gyr and at 0.8 Gyr between the first and second
    pericentres.}
  \label{gasxy}
\end{figure*}

The comparison of the SFR evolution of more gas-rich galaxies
(Fig.~\ref{gasrich1}b) with that of their less gas-rich counterparts
(Fig.~\ref{general}c) shows that long before the first close passage,
the SFR of the more gas-rich system is more active (around 30
M$_{\odot}/$yr) than that of their less gas-rich counterpart.  This is
mainly due to the higher fraction of gas available for star formation.
Previous works \citep[]{Bournaud07, Bournaud08} have suggested that
gas-rich galaxies might experience greater instability on the gaseous
disc which can lead to an intense star formation activity via
fragmentation and clump formation.  In order to check this behaviour
in our simulations, we compare the evolution of gaseous discs for
galaxies in SimI and SimVI. Fig.~\ref{gasxy} shows projected gas
distributions of one member of the interacting system in SimI (upper
panels) and in SimVI (lower panels), computed for different snapshots
during the interaction. From these figures it is clear that, while the
gaseous disc in the less gas-rich simulation evolves more smoothly, in
the more gas-rich system it has highly fragmented and formed high
density clumps where the gas satisfies the condition to form stars in
agreement with previous works\footnote{Our less gas-rich simulation
  also shows clumps along the disc but with a significant lower level
  of fragmentation than in the more gas-rich counterpart.  The
  contribution of these clumps to the SF activity in the less gas-rich
  mergers and a comparison with observations will be carried out in a
  future paper.}.  We checked that the Jeans mass is resolved by at
least 4 gas particle so that fragmentation can be numerically
described.  As it can be appreciated in the first low panel, the more
gas-rich system shows high level of disk fragmentation well before the
first close passage at 0.3 Gyr, when the two interacting galaxies are
separated by more than 100 kpc and consistent with the peak in the SFR
(see Fig.~\ref{gasrich1}b). This suggests that the internal
fragmentation of gas-rich discs is principally driven by their own
self-gravity.  To prove that this early strong SF activity is not
induced by the interaction, we run an isolated galaxy case with the
same star formation and SN parameters.  The clump formation is also
present in our isolated galaxy run.  Previous works have shown that
galaxy interactions reinforce the level of disc fragmentation for
mergers at low and high redshifts
\citep[][respectively]{teyssier10,bournaud11}.  The detailed analysis
of this issue will be part of a future paper focused on clump
formation in isolated and interacting disc galaxies.

After the intense SF activity reported during the early stage of
interaction, the SFR in the more gas-rich galaxies begins to decline
with two subsequent maxima shortly after the first and second
pericentres, respectively (Fig.~\ref{gasrich1}b).  The induced star
formation activity during the close encounters is less important than
that expected considering the strong gas inflows experienced by these
galaxies at these stages of evolution (Fig.~\ref{gasrich1}a).  We
investigate which mechanism could be operating against the triggering
action of gas inflows to inhibit the SF activity.  We analysed the
thermodynamical properties of the gas, finding that its internal
energy (dotted line of Fig.~\ref{gasrich1}b), increases continuously
fed by the SN feedback. This process is responsible of heating up and
blowing away an important fraction of gas with an increasing
efficiency as the galaxy approaches the second pericentre when it
loses $\sim 60\%$ of its initial gas mass ($\sim 10\%$ is lost at the
first pericentre and more than $80\%$ at the merging phase).  As a
consequence, the level of star formation activity is not only directly
correlated with the strength of the gas inflows but it is also
modulated by the SN feedback which is, in this simulation, more
important than in the less gas-rich interactions because of the larger
amount of available gas to feed star formation.

\begin{figure}
  \centering
  \includegraphics[width=7.cm,height=5.5cm]{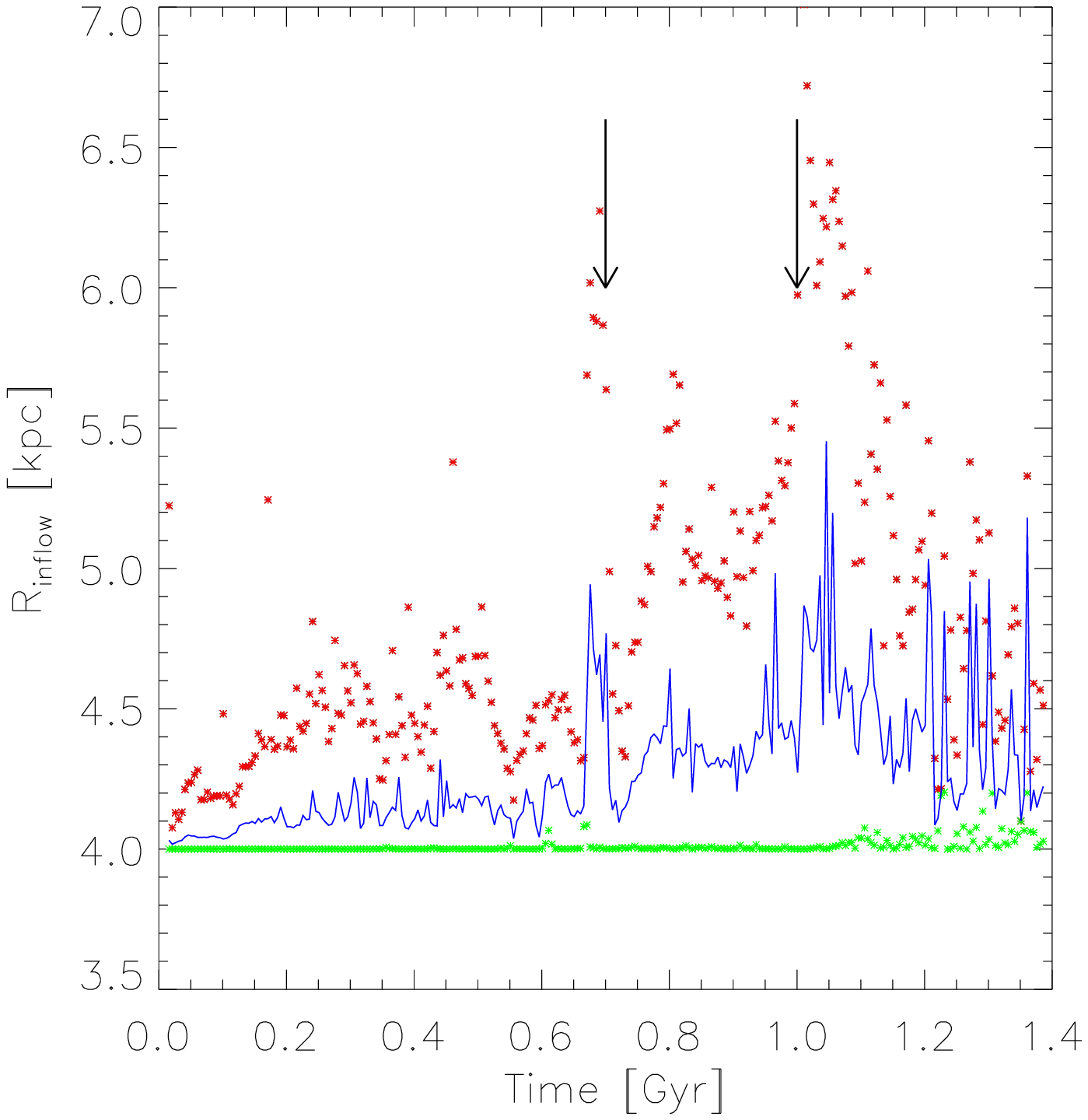}
  \includegraphics[width=7.cm,height=5.5cm]{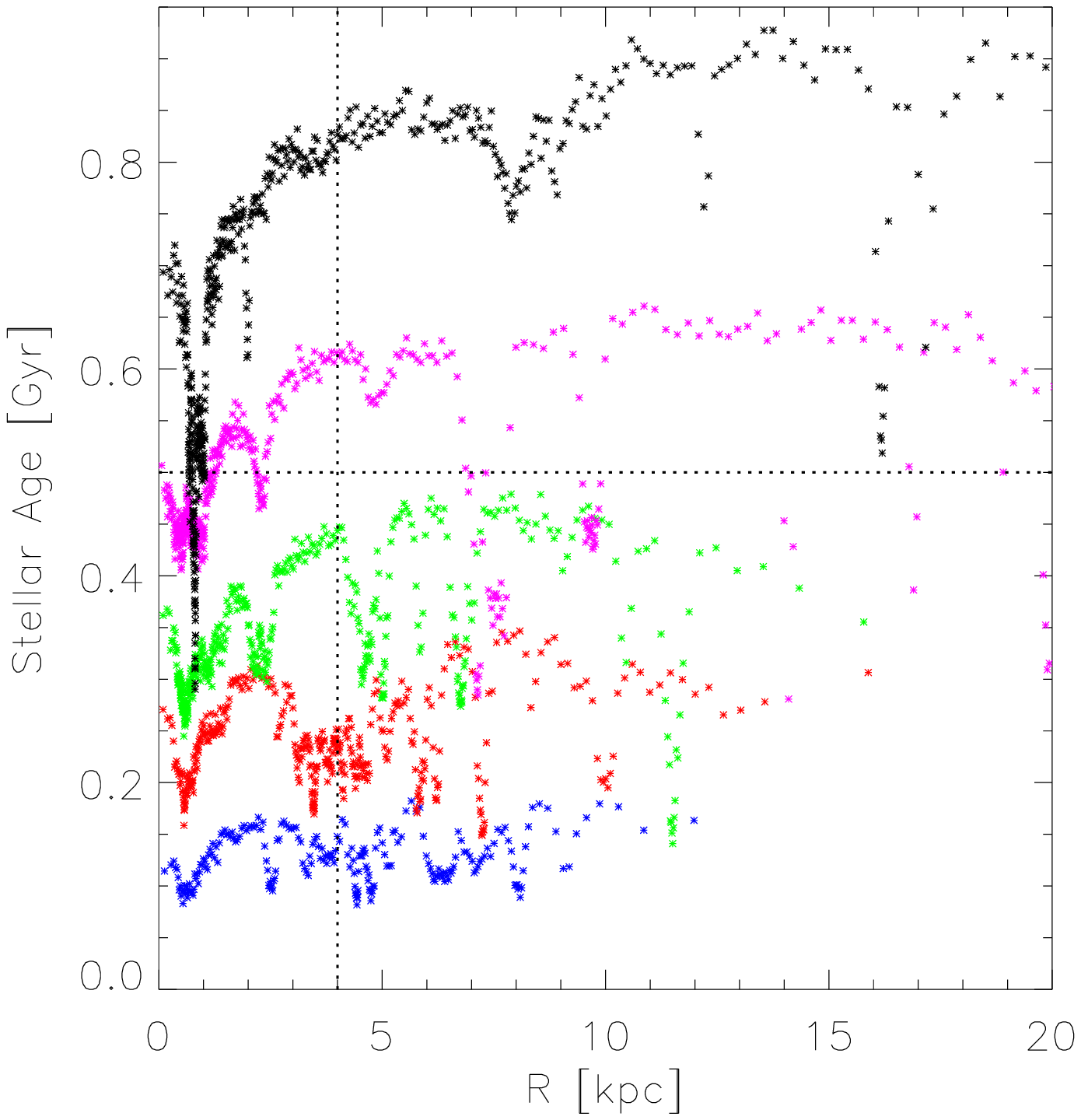}
  \caption{Upper panel: minimum (green), median (blue solid line) and
    maximum (red points) location on the disc of gas particles that
    will be immediately incorporated into the gas inflow in the
    more gas-rich simulation (SimVI). Arrows indicate the location of first
    and second pericentres. Lower panel: stellar age profiles for
    different stages of evolution of SimVI, corresponding to: 0.3 Gyr
    (blue), 0.5 Gyr (red), 0.66 Gyr (green, first pericentre), 0.86
    Gyr (magenta, after the first pericentre) and 1.16 Gyr (black,
    second pericentre).  Vertical line indicates $R_{\rm cen}$.}
  \label{edades}
\end{figure}

We investigate the chemical evolution of the ISM by analysing the O/H
abundances and $\alpha$-enhancements as in the less gas-rich interactions.
Fig.~\ref{gasrich1}c shows the evolution of oxygen abundance of the
infalling gas and of the central region.  From this figure we can see
that the central metallicity of the gas increases steadily up to the
first pericentre. We can also appreciate that the gas inflows, 
driven into the central regions during the first main starburst,
 are increasingly oxygen-rich as a consequence of 
the on-going important star formation activity.
This is a consequence of the first early starburst which is not
present in less gas-rich interactions.  However, from the first to the
second pericentre, the low metallicity of the infalling gas produces
the oxygen dilution of the central region just as in wet local 
interactions (Fig.~\ref{general}b).  The different levels of
enrichment of these gas inflows are due to their different origins:
high-metallicity inflows involved gas located nearby the central
region while lower metallicity inflows come from more external
ones. Since, our discs have initially metallicity gradients, these
different origins imply different levels of enrichment. To visualize
this, in Fig.~\ref{edades} (upper panel) we show the minimum, maximum,
and the median location on the disc plane of those gas particles which
will be immediately incorporated into the gas inflows\footnote{The
  location is determined as the distance of a given particle to the
  center of mass just before it  moves into the central region, $R <
  Ṛ_{\rm cen}$.}.  Before the first pericentre most of the gas that
falls into the central regions comes from about $4.5 $ kpc and hence,
is principally driven by internal dynamics while during the
interactions the tidally driven gas inflows are able to transport
material inward from further away regions of the disc as can be seen
in Fig.~\ref{edades} (upper panel).

In Fig.~\ref{edades} (lower panel) we show the stellar age gradients
as a function of time. We can see sharp drops in the profiles as
signatures of a discontinuous SF distribution along the disc,
reflecting the clumpy SF activity.  Profiles in blue, red, and green
show the evolution of the SF distribution before the first
pericentre. As shown by these profiles, the strong SF activity
reported during this period (Fig.~\ref{gasrich1}b) is located not only
in the central region but also extended in the outskirt of the galaxy,
contributing to enrich the ISM.  As the interaction goes on, the main
star formation activity takes place principally in the central regions
which gets rejuvenated.

As a consequence of the contribution of a low-metallicity gas inflows
during the close interaction (between the two first pericentres) the
central O/H abundance decreases by $\approx 0.4$ dex
(Fig.~\ref{gasrich1}c, dotted line).  The fact that the central oxygen
abundance decreases more slowly than that of the infalling gas, at
this stage of the interaction, is a consequence of the SF
  activity being more concentrated in the central region, continuously
  supplying new chemical elements via SNII events.  After the second
pericentre, we observe a second SFR enhancement driven by new gas
inflow (Figs.~\ref{gasrich1}a,b).  This SF activity is very
centralized in the nuclear region of the galaxy as shown in
Fig.~\ref{edades} (lower panel), producing the increase of the central
O/H abundance (Fig.~\ref{gasrich1}c, dotted line).

Finally, we also investigate the $\alpha$-enhancement of wet
encounters at high redshift.  Fig.~\ref{gasrich1}d shows that the
initial [O/Fe] value is consistent with the $\alpha$-enhancement
observed in less gas-rich interactions (Fig.~\ref{general}d).  The
subsequent evolution of the $\alpha$-enhancement shows that
immediately after the second pericentre and simultaneously to the SFR
enhancement, the central [O/Fe] (Fig.~\ref{gasrich1}d, dotted line)
exhibits a sharp increase produced by the important increase of SNII
feedback associated to the more active central SF.  After this
$\alpha$-enhancement, we find a significant drop in the central [O/Fe]
values of around 0.8 dex. This drop is caused by the injection of Fe
elements into the ISM by SNIa, whose progenitors formed $\sim$ 1 Gyr
before, during the main starbursts.

\subsection{Oxygen abundance profiles of wet interactions at high redshift}

\begin{figure}
  \centering
  \includegraphics[width=65mm]{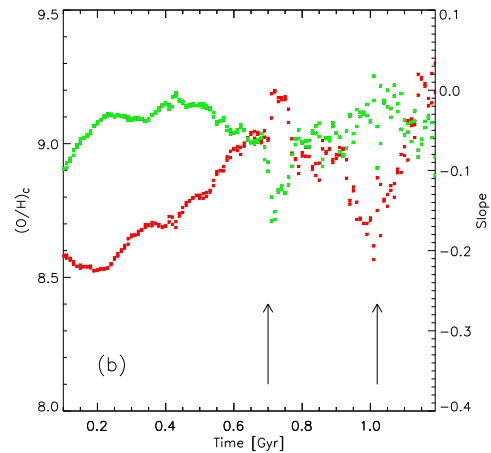}
  \caption{ Evolution of the linear regression coefficients
    ((O/H)$_{C}$: red and Slope: green) of the oxygen abundance
    profiles in the more gas-rich encounter (SimVI) computed in a range of
    3$\epsilon_{G}-7$ kpc.  Arrows indicate the first and second
    pericentres.}
  \label{perfiles}
\end{figure}

In this section, we quantify the effect of wet interactions at high
redshift on the gas-phase metallicity profiles.  Similarly to the case
of wet interactions in the local Universe, the evolution of
metallicity profiles in the more gas-rich interacting galaxies show a
clear flattening and the central dilution of O/H abundances.  We
quantify the evolution of the metallicity profiles by performing
linear regressions. Fig.~\ref{perfiles} shows the (O/H)$_{C}$ and
slopes as a function of time (red and green points, respectively).

As expected, we find that the evolution of (O/H)$_{C}$ matches that of
the central abundances computed within $R_{\rm cen}$ (see dotted line
of Fig.~\ref{gasrich1}c). We can see an increase of the (O/H)$_{C}$ in
around 0.6 dex during the approaching phase while after the first
pericentre, the decrease reflects the impact of the low-metallicity
inflows driven by the interactions.  These low-metallicity inflows
produce a decrease of $\approx 0.5$ dex in the oxygen central
abundance while in less gas-rich counterparts, the decrease is of $\approx
0.2 $ dex towards the final merging stage (Fig.~\ref{fit1}).

The evolution of the gradients shows an approximately constant slope
until the first pericentre where the profiles get more negative as a
result of the oxygen carried into the central region by the early
metal-rich inflows.  However, from the first pericentre there is a
continuous mean flattening of the metallicity profiles which gets even
slightly positive around the second pericentre where again, it starts
to be get steeper toward more negative values as more oxygen is pumped
in by SNII produced during the second tidally-induced starburst (see
Fig.~\ref{gasrich1}b).

\section{Discussion}

Several interesting aspects remain to be understood or even
acknowledged in galaxy interactions and their impact on chemical
properties.  Numerical simulations are, indeed, the most suitable tool
to study them.  Different authors with different techniques and levels
of complexity have made important contributions
\citep[e.g.][]{perez06,rupke10a,montuori10} to this area over the last
years.  All works agree to show the central abundance dilution due to
low-metallicity inflows triggered during the interaction.

\begin{figure}
  \centering
  \includegraphics[width=84mm]{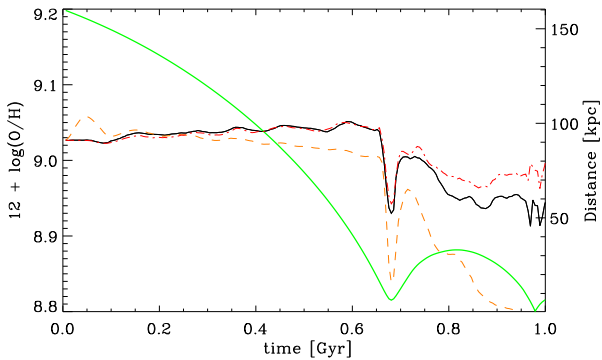}
  \caption{Effect of star formation and feedback: evolution of central
    gas-phase oxygen abundances ($ R < R_{\rm cen}$) of one galaxy in
    SimII (black full line), SimIII (orange dashed line), and Sim IV
    (red dash-dotted line). The relative distance between the centre
    of mass of the simulated galaxies is also plotted (green
    line). }
  \label{feedback}
\end{figure}

Our simulations allow us to contribute further to explore the role
played by the induced SF and SN feedback in modulating the strength of
this metallicity dilution and the evolution of the metallicity
gradients.  Particularly, we run SimIII which only follows the dynamics
of the gas with fixed metallicity assigned according to an initial
abundance profile. No new star formation activity is permitted in
SimIII.  As it can be appreciated by comparing the evolution of the
central oxygen abundances in SimII and SimIII in Fig.~\ref{feedback},
neglecting the subsequent star formation and chemical enrichment leads
to an overestimation of the metallicity dilution (and flattening of
the gradients).

One of the effects of star formation is the generation of new SN II
which contributes with new chemical elements and also pumps energy
into the ISM.  The strongest the gas inflows, the strongest the
starbursts and consequently, the impact of SN feedback which can heat
up and blow away significant fractions of the remaining enriched gas,
regulating the subsequent star formation activity and modulating the
chemical enrichment.  In SimIV we use a lower energy input for SN
event and thus, reducing the impact of energy feedback. As a
consequence, a more intense star formation can develop producing a
higher chemical enrichment than in SimII.  From Fig.~\ref{feedback} we
can see that the central oxygen abundance is very similar to that of
SimII until the first pericentre while after that, the mean oxygen
abundance in SimIV is larger since the impact of the mass-loaded
outflows are weaker and are not so efficient at transporting material
outside the central region and at inhibiting the star formation
activity (the effects on the star formation rate, $\alpha$-enhancement
and metallicity profiles can be visualized comparing results from
SimII and SimIV in Figs. ~\ref{alpha1}, ~\ref{alpha3}, and
~\ref{fit1}).  Hence, it is important to model the on-going star
formation activity in a self-consistent way  with energy and chemical SN feedback in
order to have a more comprehensive picture of the evolution of the
metallicity and gradients during galaxy interactions.

\begin{figure*}
  \centering
  \includegraphics[width=84mm]{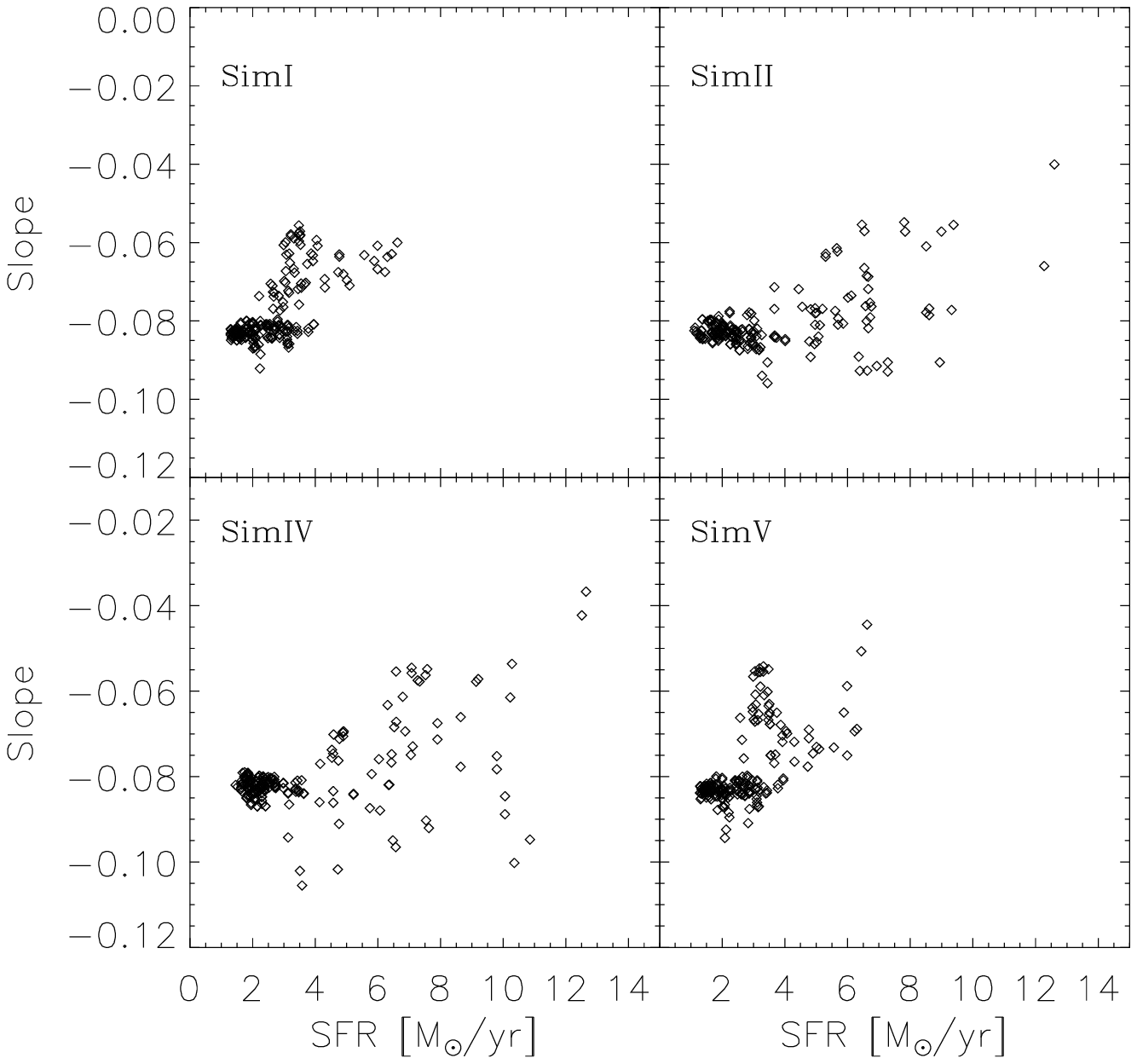}
  \includegraphics[width=84mm]{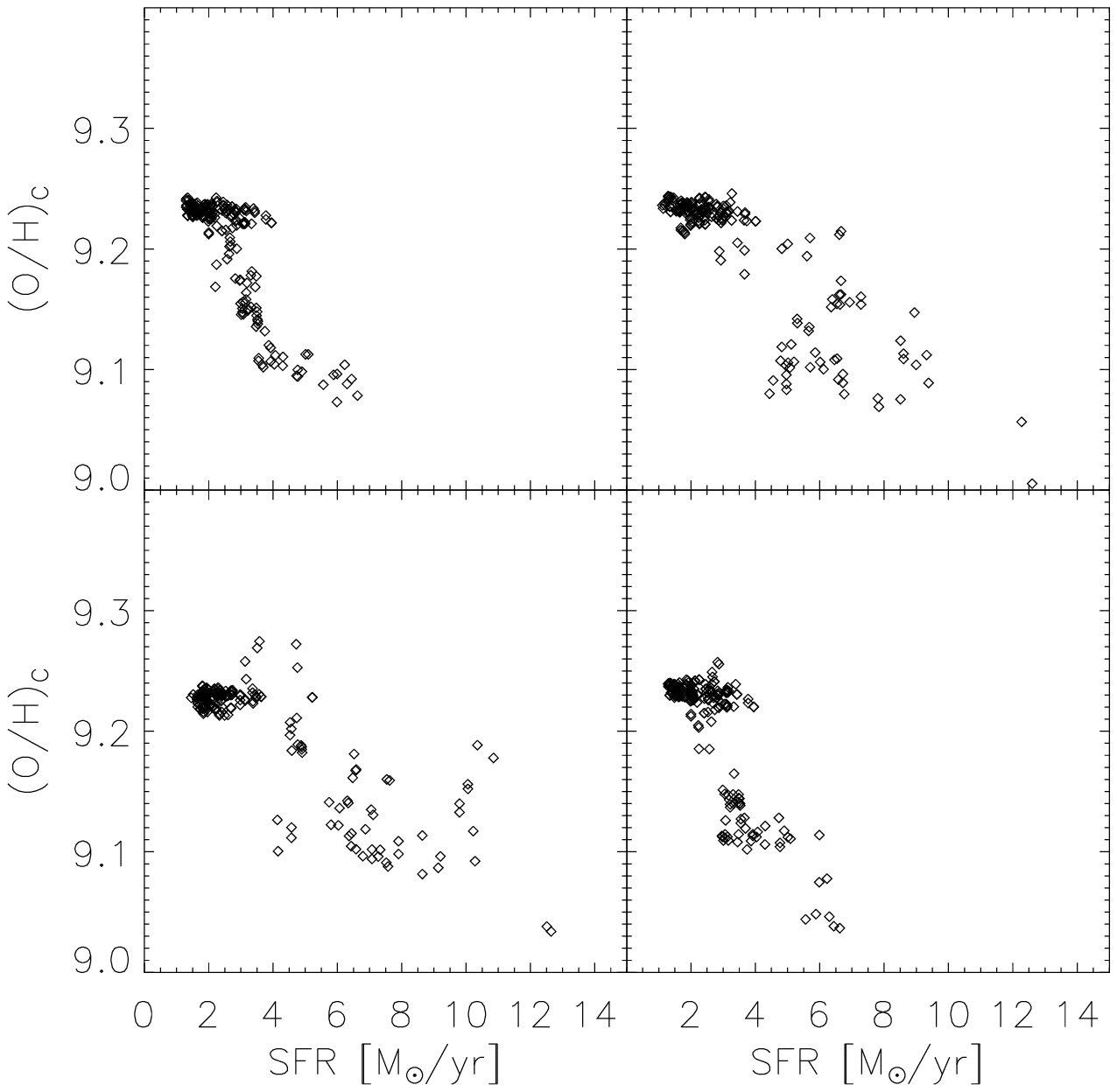}
  \caption{Metallicity gradients and central abundances as a function
    of star formation rate for interacting galaxies in SimI, SimII,
    SimIV, and SimV.}
  \label{sfroh}
\end{figure*}

As a consequence of this complex interplay of various physical
processes, we find that some a priory obvious correlations are
actually difficult to detect.  One example of this is the use of
$\alpha$-enhancement to clock the interaction as we have already
discussed in Section 3.1.  Another expected correlation is that
between the strength of the central starbursts and the metallicity
dilution.  In Fig.~\ref{sfroh} we plot the slopes of the metallicity
gradients and the central oxygen abundances as a function of star
formation rate for wet low-redshift interactions.  We can see that
globally there is a correlation so that the highest the star formation
rate, the strongest the metallicity dilution and the shallower the
abundance profiles. However, the relations depend strongly on the
orbital parameters. SimI and SimV have weaker starbursts than SimII
but nevertheless, they experienced metallicity dilution in the central
regions due to the tidally-induced gas inflows.  Hence, according to
our findings, it might be difficult to establish this correlation
observationally where galaxy pairs can be observed at different stages
of evolution and with a variety of orbital parameters.

Regarding the gas-richness, from Fig.\ref{perfiles} we can see that,
although there is a global trend in diluting central metallicity and
in flattening the mean gradients, there are also fluctuations which
correlate with the triggering of inflows and the injection of fresh
oxygen by the new SN II. Nevertheless, the mean metallicity dilution
is stronger than the less gas-rich counterpart (Fig.~\ref{fit1}). Hence, the
gas-richness is another key factor.

Depending on the astrophysical properties and orbital parameters,
galaxies in pairs can be located differently on the mean
mass-metallicity relation depending on the stage of the interaction at
which they are observed.  This effect might be particularly relevant
for the study of high redshift galaxies and could be at the origin of
the apparent increase with redshift of the scatter in the
mass-metallicity relation. It could also explain the presence of
outliers in the local mass-metallicity relation.  For example, more
gas-rich systems observed just as they are first approaching might be
dominated by the massively and clumpy star formation activity taking
place at all radii, before the interaction can actually trigger larger
scale inflows capable to transport low-metallicity gas into the
central region. Hence, these galaxies might show an excess of oxygen
\citep[e.g.][]{LMD08}.

\section{Conclusion}

By using hydrodynamical simulations of interacting pre-prepared
galaxies of comparable masses, we study the chemical evolution of the
interstellar medium during the interactions. Our simulations include
self-consistently, SN feedback and a detailed chemical model, which
allows us to follow the evolution of the gas-phase metallicity from
the initial to the final stages of an interaction. These simulations
constitute a unique set to study the effects of interactions on
chemical properties.

We analyse the different chemical response to galaxy interactions,
varying orbital configurations, strength of energy SN feedback and
initial gas fractions in the galaxy discs.  Particularly, we focused
our study on wet interactions in the local Universe. For comparison, 
we also analysed a  more gas-rich interaction in order to reproduce 
wet  mergers  typical of high redshift.

Our main results can be summarized as follows:

1) We find that a low-metallicity gas inflow developed from the first
close passage dilutes the central oxygen abundance and triggers
starbursts at the first and second pericentre.  We find that the
amplitude of the central oxygen dilution of close galaxy pairs is
fairly consistent with observations. This dilution is observed with a
large scatter for interacting systems separated less by than 30
kpc. The amplitude varies according to orbital configurations and
phases of interactions.

During the interactions the metallicity gradients get quite shallower
with fluctuations, such as sometimes they are a slightly positive
gradient.  Both the release of new chemical elements and energy by
SNII, produced during the tidally-induced starburst, modulates the
strength of the central dilution and the metallicity gradients. This
behaviour seems in good agreements with what it is observed for galaxy
pairs \citep[observations by][]{KGB06,kewley10}.  We note that it is
not possible from these simulations to test the claims of
\citet{Cresci10} since our boundary conditions are not consistent with
a cosmological scenario where gas accretion from the intergalactic
medium can fuel the galaxies.

2) We find an increase of the interstellar [O/Fe] values, observed
after the first pericentre as suggested by previous works.  This
$\alpha$-enhancement does not only reflect the tidally-induced star
formation activity but is also affected by galactic outflows and the
abundances of the gas inflows.  As a consequence, we show that the
$\alpha$-enhancement can only indicate the proximity to the first
pericentre and in those encounters with non coplanar retrograde
orbits.  After the first pericentre, gas inflows might dilute the
central [O/Fe] values, erasing the trace left by the star formation
activity, and restarting the clock to date the interaction.  We also
notice that the period of time where this enhancement could be
observed is quite short $(< 0.2$ Gyr).

3) More gas-rich interactions show massive and clumpy star formation
on the disc from early stages of the interaction.  During the
interaction, from the first pericentre to the final stages, we find
similar behaviours as in the case of the less gas-rich runs with the
particularities imprinted by the more gas availability.  Gas-richness
has an impact on the strength of low-metallicity gas inflows which in
the simulated interaction, are able to dilute the central abundances
by $ 0.5$ dex, a factor of 2.5 larger than in its less gas-rich
counterpart.  There are important variations in the central abundances
and gradients as the interaction of more gas-rich galaxies proceeds
due to the competition between various processes, such as the
injection of new chemical elements after starbursts, the action of
outflows which blow away part of the enriched material, and the gas
dynamics of clumpy discs.

4) We find a correlations between the metallicity gradient, the
central chemical abundance and the strength of the star formation
activity.  However, the slope and scatter of these correlations depend
on orbital parameters, SN outflows and gas-richness. Hence, it is not
a simple relation which can be easily unveiled observationally.  The
metallicity gradients are affected by the strength of the central
nuclear starbursts and the extended star formation detected in the
outer regions of the galaxies. Star formation in clumps are also found
to play a role in modulating the metallicity gradients.  A detailed
analysis of disc fragmentation as a function of gas richness will be
carried out in a forthcoming paper.

Our results suggest that metallicity evolution is closely related to
gas dynamics during the encounters. Denying processes that affects the
gas dynamics and its level of enrichment such as star formation and SN
feedback, can contribute either to overestimate or underestimate the
effects on the central abundance dilution or metallicity gradients.
These findings show that there are still many issues to be addressed
in galaxy interactions and their role in chemical evolution.

\section*{Acknowledgments}
We thank the referee, F. Bournaud, for a constructive report.  PBT
thanks ISIMA 2010 for the hospitality and challenging environment
which allowed part of this work to be done. This work was partially
supported by the Consejo Nacional de Investigaciones Cient\'{\i}ficas
y T\'ecnicas (PIP 2009/0305), Agencia Nacional de Promoci\'on
Cient\'\i fica y T\'ecnica (PICT 32342 (2005)), Max Planck 245 (2006),
the Agence Nationale de la Recherche (ANR-08-BLAN-0274-01).

\label{lastpage}
\end{document}